\documentclass[a4paper,11pt]{article}
 \pdfoutput=1 % if your are submitting a pdflatex (i.e. if you have
 \usepackage{jheppub} % for details on the use of the package, please
 \usepackage[T1]{fontenc} % if needed
 \usepackage{footnote}

\title{\boldmath
Induced gravitational waves from the cosmic coincidence
}
\author[1,2]{Shyam Balaji,}
\author[2, 3, 4]{Joseph Silk,}
\author[1]{and Yi-Peng Wu}

\affiliation[1]{Laboratoire de Physique Th\'{e}orique et Hautes Energies (LPTHE), \\
	UMR 7589 CNRS \& Sorbonne Universit\'{e}, 4 Place Jussieu, F-75252, Paris, France}
\affiliation[2]{Institut d’Astrophysique de Paris, UMR 7095 CNRS \& Sorbonne Universit\'{e}, 98 bis boulevard Arago, F-75014 Paris, France}
\affiliation[3]{Department of Physics and Astronomy, The Johns Hopkins University, 3400 N. Charles	Street, Baltimore, MD 21218, U.S.A.}
\affiliation[4]{Beecroft Institute for Particle Astrophysics and Cosmology, University of Oxford, Keble	Road, Oxford OX1 3RH, U.K.}

\emailAdd{sbalaji@lpthe.jussieu.fr}
\emailAdd{silk@iap.fr}
\emailAdd{ywu@lpthe.jussieu.fr}

\abstract{
The induced gravitational wave (GW) background from enhanced primordial scalar perturbations is one of the most promising observational consequences of primordial black hole (PBH) formation from inflation. We investigate the induced GW spectrum $\Omega_{\textrm{IGW}}$ from single-field inflation in the general ultra-slow-roll (USR) framework, restricting the peak frequency band to be inside $10^{-3}$-$1$ Hz and saturating PBH abundance to comprise all dark matter (DM) in the ultralight asteroid-mass window. By invoking successful baryogenesis driven by USR inflation, we verify the viable parameter space for the specific density ratio between baryons and PBH DM observed today, the so-called ``cosmic coincidence.'' We show that the cosmic coincidence requirement bounds the spectral index $n_{\rm UV}$ in the high frequency limit, $\Omega_{\textrm{IGW}}(f\gg 1)\propto f^{-2n_{\rm UV}}$, into $0 < n_{\rm UV} < 1$, which implies that baryogenesis triggered by USR inflation for PBHs in the mass range of $10^{-16}$-$10^{-12} M_\odot$ can be tested by upcoming Advanced LIGO and Virgo data and next generation experiments such as LISA, Einstein Telescope, TianQin and DECIGO.   
}

\begin{document}

\maketitle
\flushbottom

\section{Introduction}
%%% Introduction for induced GWs.
The cosmological gravitational wave (GW) background induced by scalar-type density fluctuations at non-linear orders must exists in all viable scenarios for the early universe \cite{10.1143/PTP.37.831,Matarrese:1993zf,Matarrese:1997ay,Ananda:2006af,Baumann:2007zm,Martineau:2007dj,Bartolo:2007vp,Saito:2008jc,10.1143/PTP.123.867,PhysRevD.81.023517,Bugaev2010,PhysRevD.83.083521,Suyama:2011pu,Assadullahi:2009jc,Assadullahi:2009nf,Arroja:2009sh,Alabidi:2012ex,Alabidi:2013lya,Kawasaki:2013xsa} (see \cite{Domenech:2021ztg} for a recent review). In most of the scenarios, such as inflationary models or alternatives, the induced GWs generated by primordial scalar perturbations at second order are already too small and thus very challenging to observe \cite{Ananda:2006af,Baumann:2007zm,Martineau:2007dj,Bartolo:2007vp}, yet exciting opportunities still exist 
%in the frequency band of $10^{-4}$--$10^3$ Hz 
for future space-based telescopes that may search for largely enhanced GW spectra that are closely related to primordial black hole (PBH) formation \cite{Saito:2008jc,10.1143/PTP.123.867,PhysRevD.81.023517,Belotsky:2014kca,Bugaev2010,PhysRevD.83.083521,Suyama:2011pu,Suyama:2014vga,Nakama:2015nea,Nakama:2016enz}.   

Since the first detection of GW events from a binary-black-hole merger in 2016 \cite{PhysRevLett.116.061102}, PBHs have turned into compelling dark matter (DM) candidates \cite{Carr:2016drx} and the mass windows allowed for PBHs to comprise all DM have been severely constrained by observations (see \cite{Carr:2020gox,Carr:2020xqk,Green:2020jor} for recent reviews). Given that the largely enhanced scalar power spectrum on small scales from inflation manifests the mainstream scenario for PBH formation, the induced GW background from these inflationary models is considered one of the most promising signatures of PBH DM, flourishing many intensive investigations \cite{Kohri:2018awv,Liu:2020oqe,Cai:2019cdl,Cai:2019elf,Yuan:2019wwo,Pi:2020otn,Cai:2018dig,Garcia-Bellido:2017aan,Fumagalli:2020nvq,Fumagalli:2021cel,Unal:2018yaa,Ragavendra:2020vud,Ozsoy:2020kat,Bartolo:2018rku,Tada:2019amh,Bartolo:2018evs,Ballesteros:2020qam,Wang:2019kaf,Kawai:2021edk}. Currently, the viable window for PBH DM in the ultralight asteroid-mass range seems to put the peak amplitude of the induced GW spectrum inside the joint frequency band ($10^{-4}$--$10^3$ Hz) for LIGO and the next-generation experiments \cite{Ragavendra:2020sop}, such as Einstein Telescope (ET) \cite{Maggiore:2019uih}, LISA \cite{amaroseoane2017laser,Barausse:2020rsu} and DECIGO \cite{Yagi:2011wg,Kawamura:2020pcg}.

However, if we suppose that PBHs indeed occupy a significant fraction of the DM density in today's universe, they will exhibit a surprisingly similar amount of energy density relative to that of the baryons and thus they are inevitably confronted by the so-called ``cosmic coincidence problem,''  as is the case for all other DM candidates. The cosmic coincidence problem can provide a good motivation to consider PBH DM as a consequence of inflation, due to the recent theoretical validation of successful baryogenesis triggered by single-field inflationary models for PBH formation \cite{Wu:2021gtd,Wu:2021mwy}.

In this work, we investigate baryogenesis from single-field inflation with an enhanced power spectrum for curvature perturbations on small scales driven by the ultra-slow-roll (USR) transition of the rolling dynamics of the inflaton field. We consider baryon asymmetry created by the Affleck-Dine (AD) mechanism, but we relax the constant-mass assumption for the AD field used in Refs.~\cite{Wu:2021gtd,Wu:2021mwy} so that it is possible to adopt a more general inflationary power spectrum away from the exact USR limit. This is to say that the inflationary spectrum can have a very sharp power-law decay on small scales (from its peak amplitude), which is favourable for PBH formation with a monochromatic mass spectrum. We review the general USR spectrum in Section~\ref{Sec_USR_inflation}.

After solving the generalised coherent motion of the AD field during inflation, we obtain the initial conditions for computing the final baryon asymmetry in the radiation dominated epoch, as provided in Section~\ref{Sec_B_from_USR}. In Section~\ref{Sec_CC}, we explore the modified parameter space from successful baryogenesis through general USR inflation and clarify its indication to the cosmic coincidence problem (Section~\ref{sec:cosmiccoincidence}). We compute the spectrum of the induced GWs associated with the input USR inflation and we discuss prospects for the validity of the ``cosmic coincidence'' residing in the PBH DM paradigm and the implications for current and future experiments. For completeness, we address in Section~\ref{Sec_non_G} the corrections to the induced GW spectrum led by non-Gaussianity of the curvature perturbations and we briefly illustrate the GW counterpart that result from the binary PBHs mergers in this scenario. Finally, our conclusions are provided in Section~\ref{Sec_conclusion}.

%%% Introduction for baryogenesis from USR inflation.

\section{Generalised ultra-slow-roll inflation}\label{Sec_USR_inflation}
\begin{figure}
	\begin{center}
		\includegraphics[width=10cm]{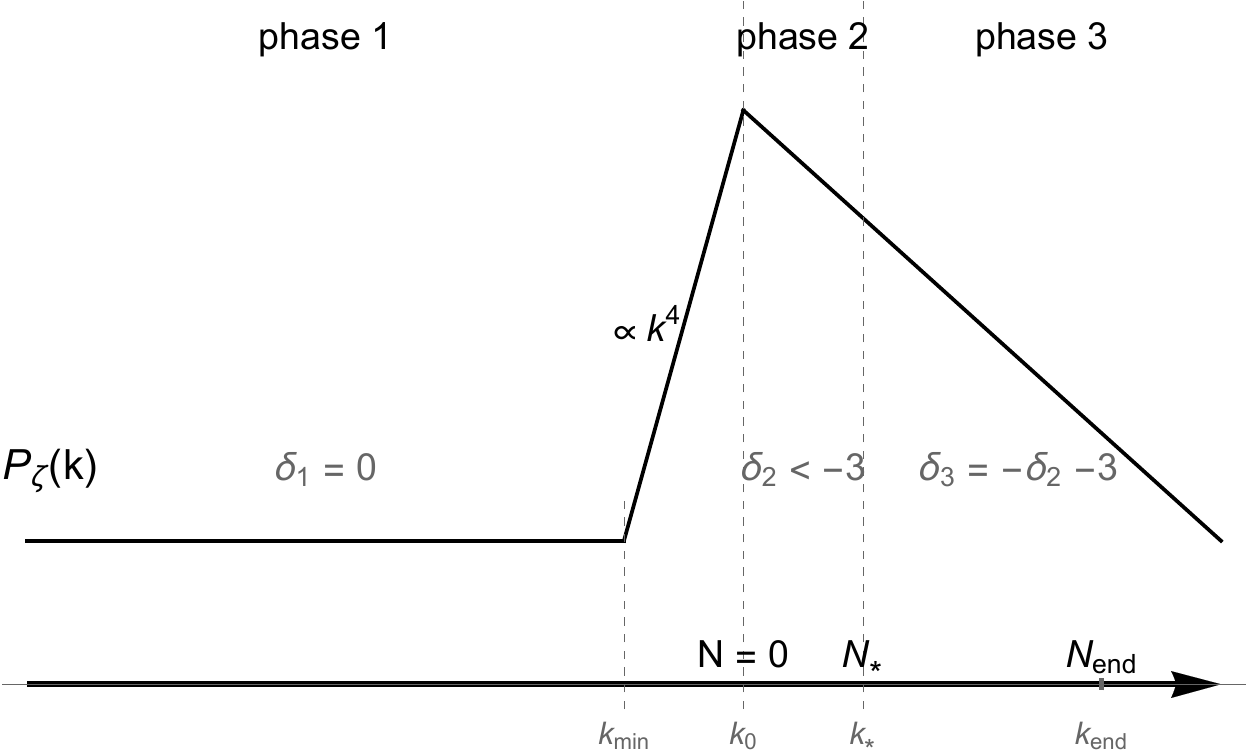} 
		%\par
	\end{center}
	\caption{The broken power-law template for the power spectrum of the curvature perturbation $P_{\zeta}$ based on a 3-stage inflation. The rate-of-rolling $\delta_1 = 0$ is used in the primary slow-roll phase (phase 1) and $\delta_2 < -3$ in the transient ultra-slow-roll (USR) phase (phase 2) is allowed to be away from the exact ultra-slow-roll limit ($\delta_2 = -3$). The post-USR phase (phase 3) exhibits a continuous scaling from phase 2 yet it must be an acceleration phase with $\delta_3 = -\delta_2 -3 >0$ to terminate inflation. \label{fig:USR_template}}
\end{figure}

In this work, we consider PBH formation from single-field models of inflation that experience a transient USR phase with largely enhanced curvature perturbation $\zeta$ on very small scales. The pivot scale $k_0$ that denotes the peak position of the power spectrum $P_\zeta(k)$ should be in the range of $k_0 \sim 10^{12} - 10^{15}$ Mpc$^{-1}$ so that the resulting PBHs formed during radiation domination are inside the ultra-light asteroid-mass window $M_{\rm PBH}/M_\odot \sim 10^{-16} - 10^{-12} $ and they are still allowed to account for all dark matter. The three most relevant parameters that one should keep in mind throughout this work, are defined as
\begin{align}\label{def_USR_parameters}
\delta \equiv \frac{\ddot{\phi}}{H\dot{\phi}},\qquad \Delta N \equiv N_\ast - N_0,\qquad N_{\rm end},
\end{align}
where $\delta$ parametrises the rate of rolling of inflaton $\phi$, $\Delta N$ measures the duration of the USR phase and $N_{\rm end}$ is the e-fold number at the end of inflation. For convenience, we set the e-fold number $N_0 = 0$ at which the pivot scale $k_0$ crosses the horizon.

As shown in Figure~\ref{fig:USR_template}, the generic scaling behavior of the power spectrum $P_\zeta$ as a function of $k$ can be summarized by a template of the broken power-law form \cite{Ballesteros:2020sre,Byrnes:2018txb,Carrilho:2019oqg,Cheng:2018qof,Liu:2020oqe,Ng:2021hll,Ozsoy:2019lyy}
\begin{eqnarray}\label{USR_template}
P_\zeta(k) = \left\{
\begin{array}{ll}
A_{\rm CMB} &\; k < k_{\rm min}, \\
A_{\rm PBH} (k/k_0)^4, &\; k_{\rm min} < k < k_0, \\
A_{\rm PBH} (k/k_0)^{6 + 2\delta_2}, &\; k > k_0,
\end{array}
\right.
\qquad
A_{\rm PBH} =  A_{\rm CMB} \left(\frac{k_{0}}{k_{\ast}}\right)^{6 + 4 \delta_2}.
\end{eqnarray}
Where $A_{\rm CMB} = \frac{H_\ast^2}{\epsilon_{\rm CMB} M_P^2}\approx 2.2\times 10^{-9}$ is measured by CMB experiments and $k_\ast$ is the comoving scale that crosses the horizon at $N = N_\ast$ (the end of the transient USR phase). The duration of the USR phase is $\Delta N = \ln \frac{a_\ast}{a_0} = \ln \frac{k_\ast}{k_0}$. 
%and $A_{\rm PBH}/A_{\rm CMB} = e^{-N_\ast(6+4\delta_2)}$.

The broken power-law template given by \eqref{USR_template} includes three phases of inflation, which are denoted as the primary slow-roll phase (phase 1), the transient USR phase (phase 2) for enhancing the amplitude of the power spectrum and the post-USR phase (phase 3) with $\delta > 0$ that can terminate inflation. It is remarkable that the curvature perturbation $\zeta_k$ in the range of $k_{\rm min} \leq k < k_0$ are modes that have exited the horizon in phase 1, where $k_{\rm min} \approx k_0\left(\frac{k_0}{k_\ast}\right)^{-3/2-\delta_2}$. These modes undergo superhorizon evolution after the USR transition into phase 2 \cite{Cheng:2018qof}, and eventually settle in the final boundary measured at the end of inflation as a $P_\zeta \sim k^4$ growth \cite{Byrnes:2018txb,Carrilho:2019oqg,Leach:2001zf}. The $P_\zeta \sim k^4$ growth is a consequence of entropy perturbation domination \cite{Leach:2000yw}, which is the criterion for breaking the initial scaling power $P_\zeta \sim k^0$ led by phase 1 \cite{Leach:2001zf,Ng:2021hll}.\footnote{The domination of entropy mode in $\zeta$ is so far a sufficient condition for violating the continuity of the momentum scaling $\Delta$ in the power spectrum as $P_\zeta \sim k^{2\Delta}$ is protected by the dilatation symmetry of the de Sitter background. Note that $\Delta = 3/2-\vert 3/2 +\delta\vert \rightarrow 0$ in the slow-roll and exact USR inflation with $\delta =0$ or $\delta = -3$, respectively. It is interesting to seek other mechanisms that can break the scaling power $\Delta$ in single-field inflation.}
 On the other hand, the entropy domination does not occur in most of the single-field models across the transition from phase 2 to 3 so that $P_\zeta(k)$ exhibits a continuous scaling in the power of $k$ till the end of inflation. This adiabatic condition fixes the rate-of-rolling in phase 3 as $\delta_3 = -\delta_2 -3$ \cite{Ng:2021hll}.  

The previous studies \cite{Wu:2021mwy,Wu:2021gtd} considered a transient quasi-USR phase with $ -3.2 \lesssim \delta_2 < -3$ to sustain the constant-mass approximation for the charged scalar that is responsible for creating the baryon asymmetry. In this work, we relax such a restriction by considering a general value allowing for $\delta_2 \ll -3$.
\footnote{In single-field inflation, the rolling rate $\delta < -3$ is generally realized from inflaton $\phi$ to climb up a little bump in the potential $V(\phi)$ that exhibits a tachyonic mass ($\partial_\phi^2V < 0$ and $\partial_\phi V >0$) region. A transient off-attractor phase is usually required for the inflaton to obtain enough kinetic energy to climb up the bump. Such an off-attractor phase with increasing kinetic energy results in a generic dip feature in the power spectrum right before the $k^4$ growth. However, this dip feature in the power spectrum has little effect on the PBH abundance or the baryogenesis process considered in this work, so it can be removed by considering the broken power-law template \eqref{USR_template} for simplicity. As reflected from the discussion in Section~\ref{Sec_CC}, our scenario in fact focuses on the regime of $-4 < \delta < -3$ where the possible kinetic energy domination (with $\epsilon_H\sim \mathcal{O}(1)$) due to the off-attractor phase is negligibly short.} 
 There can be at least two benefits for constructing the scenario in the region of $\delta_2 \ll -3$. First, the effective mass of the inflaton in phases 2 and 3 reads $\frac{m_\phi^2}{H_\ast^2} \approx -\delta(\delta + 3)$ \cite{Ng:2021hll,Wu:2021mwy} (assuming constant $\delta$), where $m_\phi$ is continuous across the two phases as protected by the scaling condition $\delta_3 = -\delta_2 -3$. Taking $\delta_2 \ll -3$ thus gives $m_\phi/H_\ast \gg 1$, which is expected to significantly reduce the effect of quantum diffusion in the exact USR case with $\delta_2 = -3$ \cite{Ezquiaga:2019ftu,Figueroa:2020jkf,Pattison:2021oen}. Second, $P_\zeta(k> k_0) \sim k^{6+2\delta_2}$ with $\delta_2 \ll -3$ decays sharply in the large $k$ limit, leading to a narrow peak at $k = k_0$. As shown in Section~\ref{Sec_PBH_DM}, such a narrow-peak spectrum can transfer into a narrow distributed mass function for PBHs, which is more appropriate (although still not quite accurate) when comparing with most of the observational constraints in which the monochromatic PBH mass assumption is generically applied.

\section{Baryogenesis triggered by the ultra--slow--roll transition}\label{Sec_B_from_USR}
The breakdown of the constant mass approximation \cite{Wu:2021mwy} for the charged scalar is the price to pay for entering into the $\delta_2 \ll -3$ regime. In this section, we investigate the coherent motion of the charged scalar with time-varying mass induced by the transient USR transition. Given that we consider the charged scalar, $\sigma$, as the source field for generating baryon asymmetry via the AD mechanism, we also identify $\sigma$ as the AD field.

As introduced in Ref.~\cite{Wu:2021mwy}, we consider $\sigma$ possessing a $U(1)$ baryon number given by $n_B = j^0 = i(\sigma^\ast\dot{\sigma} - \sigma\dot{\sigma}^\ast)$. The dynamics of inflaton $\phi$ enters the effective mass of $\sigma$ through derivative couplings described in the Lagrangian 
\begin{align}\label{Lagrangian_AD}
\mathcal{L}_\sigma = \vert \partial\sigma\vert^2 + m_\sigma^2 \vert \sigma\vert^2 + 
\frac{c_1}{\Lambda} \left\vert  \sigma^2\right\vert \square \phi 
+ \frac{c_2}{\Lambda}\partial_\mu \phi \left[\sigma\partial^\mu \sigma + \sigma^\ast\partial^\mu \sigma^\ast\right] + 
\frac{c_3}{\Lambda^2} \left(\partial\phi\right)^2\vert\sigma\vert^2,
\end{align}
where $c_1$, $c_2$, $c_3$ are $\mathcal{O}(1)$ coupling constants. The Lagrangian \eqref{Lagrangian_AD} is C/CP invariant but the $c_2$ term violates baryon number. Note that the imaginary phase in $c_2$ can be absorbed into the phase term of $\sigma$. The cutoff scale $\Lambda$ should be in the range of $H_\ast \ll \Lambda \leq M_\textrm{P}$ to justify the effective field theory description during inflation.

To study the coherent motion of the AD field, it is more convenient to decompose the Lagrangian into the mass eigenstates $\sigma_{\pm}$ via $\sigma \equiv \frac{\sigma_{-} + i \sigma_{+}}{\sqrt{2}}$, where \eqref{Lagrangian_AD} becomes a system of two real scalars as
\begin{align}\label{Lagrangian_shift_pm_decouple}
\mathcal{L}_{\pm} =& \frac{1}{2}\left(\partial\sigma_{+}\right)^2 + \frac{1}{2} 
\left[m_\sigma^2 + \frac{c_1 + c_2}{\Lambda} \square \phi  + \frac{c_3}{\Lambda^2}\left(\partial\phi\right)^2\right] \sigma_{+}^2
\nonumber\\
&+ \frac{1}{2}\left(\partial\sigma_{-}\right)^2 + \frac{1}{2} 
\left[m_\sigma^2 + \frac{c_1 - c_2}{\Lambda} \square \phi  + \frac{c_3}{\Lambda^2}\left(\partial\phi\right)^2\right] \sigma_{-}^2 .
\end{align}
The effective masses are therefore controlled by the coherent motion of $\phi$, where
\begin{align}\label{mass_pm}
m_\pm^2 = m_\sigma^2 + \frac{c_1 \pm c_2}{\Lambda} \square\phi + \frac{c_3}{\Lambda^2} \left(\partial\phi\right)^2.
\end{align}
We note that the necessary CP violation for successful baryogenesis is spontaneously realised by the emergence of a CP-violating initial VEV in a local universe \cite{Wu:2020pej,Hook:2015foa} (similar to the idea of spontaneous T violation \cite{Lee:1973iz}).

\subsection{Time evolving scalar masses}
Let us now apply the inflationary background with the transient USR transition introduced in Section~\ref{Sec_USR_inflation}. Based on the definitions in Eq.~\eqref{def_USR_parameters}, we get the equations of motion 
\begin{align}
    \square\phi = -\ddot{\phi} -3H\dot{\phi} \approx - (\delta + 3)\sqrt{2\epsilon_H} M_PH_\ast^2
\end{align}
and 
\begin{align}
 (\partial\phi)^2 = \dot{\phi}^2 \approx 2\epsilon_HM_P^2 H_\ast^2,   
\end{align}
where $\epsilon_H = -\frac{\dot{H}}{H^2}$ is the first slow-roll parameter. Therefore, with the transition of $\delta$ in each phase of inflation, the effective masses take different values like 
\begin{align}\label{mass_i_pm}
m_{i\pm}^2 = m_\sigma^2 + \frac{c_1 \pm c_2}{\Lambda}[-(\delta_i + 3)]\sqrt{2\epsilon_i} M_PH_\ast^2
+ \frac{c_3}{\Lambda^2} 2\epsilon_i M_P^2H_\ast^2.
\end{align}
Note that $\frac{d\ln\epsilon_H}{dN} \approx 2\delta$ and $\epsilon_1 \equiv \epsilon_H(N < 0) = \epsilon_{\rm CMB}$ is approximately a constant. This yields $\epsilon_2 \equiv \epsilon_{\rm CMB}e^{2\delta_2 N}$ and $\epsilon_3 = \epsilon_\ast e^{2\delta_3(N-N_\ast)}$ where $\epsilon_\ast \equiv \epsilon_H(N_\ast) = \epsilon_{\rm CMB}e^{2\delta_2 N_\ast}$.

\begin{figure}
	\begin{center}
		\includegraphics[width=7.2cm]{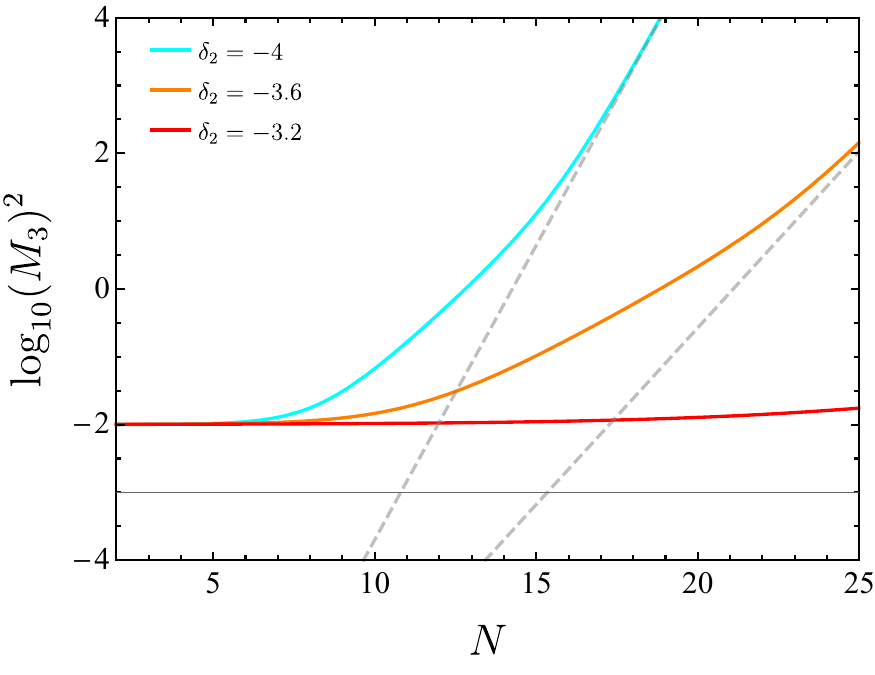} 
		\hfill
	    \includegraphics[width=7.4cm]{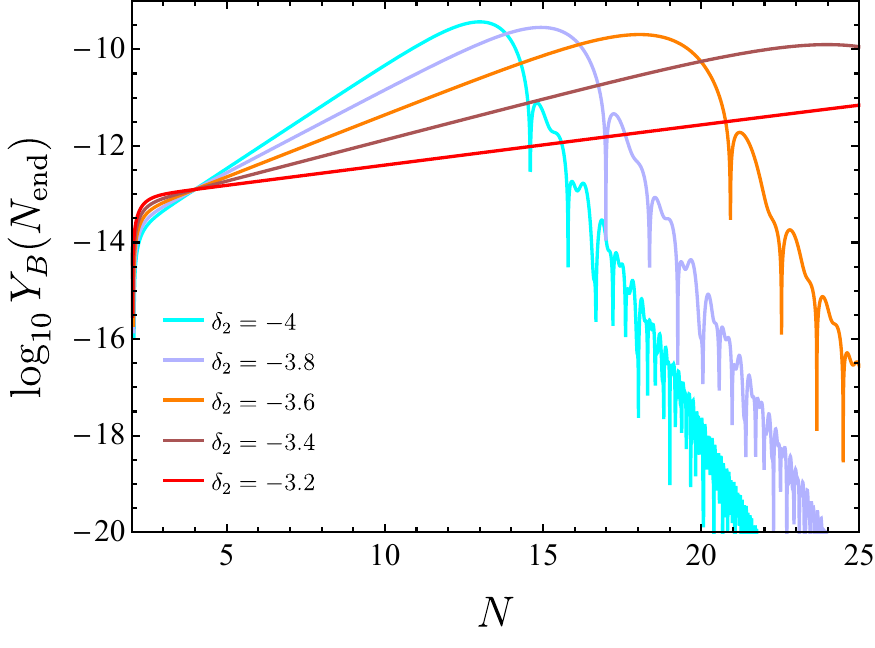} 
		%\par
	\end{center}
	\caption{ The time evolution of $M_3^2(N)$ (left panel)  and the baryon asymmetry at the end of inflation $Y_B(N_{\rm end})$ (right panel) with various choices of $\delta_2$ and $\delta_3 = -\delta_2 -3$. In both panels, $\frac{m_\sigma}{H_\ast} = 0.1$, $\frac{\Lambda}{M_P} = 1$, $\Delta N = N_\ast = 2$ are used.  \label{fig:M_3}}
\end{figure}

In order to address the time evolution of $\epsilon_i$ in the effective masses \eqref{mass_i_pm}, we use the dimensionless parametrisation
\begin{align}\label{def_M_i}
M_{i\pm}^2 \equiv \frac{m_{i\pm}^2}{H_\ast^2}
= M_\sigma^2 + A_{i\pm} e^{\delta_i N} + B_i e^{2\delta_i N},
\end{align}
where $M_\sigma = m_\sigma/H_\ast$. $A_{i\pm}$ and $B_i$ are dimensionless constants, and the coherent motion of the mass eigenstates in each phase, $\sigma_{i\pm}$ are governed by the equation of motion
\begin{align}\label{eom_sigma_general}
\frac{d^2\sigma_{i\pm}}{dN^2} + 3 \frac{d\sigma_{i\pm}}{d N} + M_{i\pm}^2(N) \sigma_{i\pm} = 0.
\end{align}
The dimensionless constants are solved to be
\begin{align}
    A_{1,2\pm} &= \frac{(c_1\pm c_2)[-(\delta_{1,2} + 3)]\sqrt{2\epsilon_{\rm CMB}}M_P}{\Lambda} ,&
    B_1 = B_2= \frac{2 c_3 \epsilon_{\rm CMB} M_P^2}{\Lambda^2}.
\end{align}
%\begin{align}
%A_{1,2\pm} &= (c_1\pm c_2)[-(\delta_{1,2} + 3)]\sqrt{2\epsilon_{\rm CMB}} \frac{M_P}{\Lambda}, \\\nonumber
%B_1 = B_2 &= 2 c_3 \epsilon_{\rm CMB} \frac{M_P^2}{\Lambda^2}.
%\end{align}
Since $\delta_1\rightarrow 0$ in phase 1, one can easily find that $M_{1\pm}$ are constants so that we may adopt the initial conditions from \cite{Wu:2021mwy} as
\begin{align}\label{initial_condition_Phase1}
\sigma_{1\pm}(N_0) = \sqrt{\frac{3}{8\pi^2}} \frac{H_\ast^2}{m_{1\pm}}, \quad \dot{\sigma}_{1\pm}(N_0) = 0,
\end{align}  
where $c_1$, $c_2$ are chosen such that $m_{1\pm}^2$ are positive.

The general solution of \eqref{eom_sigma_general} is very complicated. Fortunately, the duration of the transient USR phase (phase 2) is constrained by the PBH abundance and in fact $\Delta N = N_\ast < 3$ in all physical models with an arbitrary choice of statistical method. We have checked that $M_{2\pm}$ being approximately constant holds in the whole parameter space of interest for $N_\ast \leq 3$, and this allow us to import the constant mass results from Ref.~\cite{Wu:2021mwy}, where at the of end phase 2 we have
\begin{align}\label{sol_sigma_2}
\sigma_{2\pm} (N_\ast) &= \sqrt{\frac{3}{8\pi^2}} \frac{H_\ast^2}{m_{1\pm}} \frac{1}{2 \nu_{2\pm}}
\left( \Delta_{2\pm}^+ e^{-\Delta_{2\pm}^- N_\ast } - \Delta_{2\pm}^- e^{-\Delta_{2\pm}^+ N_\ast } \right), 
\\\label{sol_dsigma_2}
\left. \frac{d\sigma_{2\pm}}{d N}\right\vert_{N = N_\ast}  &= \sqrt{\frac{3}{8\pi^2}} \frac{m_{2\pm}^2}{m_{1\pm}} \frac{1}{2\nu_{2\pm}}
\left( e^{-\Delta_{2\pm}^+ N_\ast } -  e^{-\Delta_{2\pm}^- N_\ast }\right),
\end{align}
with definitions $\nu_{2\pm} = \sqrt{\frac{9}{4} - M_{2\pm}^2}$ and $\Delta_{2\pm}^\pm = \frac{3}{2} \pm \nu_{2\pm}$.

Now only the solutions for phase 3 remain. In phase 3, we restrict ourselves to the broken power-law template \eqref{USR_template} with $\delta_3 = - \delta_2 - 3$, where 
\begin{align}
\nonumber
A_{3\pm} &= (c_1\pm c_2)[-(\delta_{3} + 3)]\sqrt{2\epsilon_{\rm CMB}}\frac{M_P}{\Lambda} e^{(\delta_2 - \delta_{3})N_\ast}, 
\\\nonumber
B_3 &= 2 c_3  \epsilon_{\rm CMB} \frac{M_P^2}{\Lambda^2} e^{2(\delta_2 - \delta_3)N_\ast}.
\end{align}

Some examples for the time evolution of $M_3^2$ with respect to the e-fold number $N$ are given in Figure~\ref{fig:M_3}. One can see that the $B_3$ term in Eq.~\eqref{def_M_i} (illustrated by the dashed line in Figure~\ref{fig:M_3}) only comes to dominate the effective mass with sufficiently long duration in phase 3. Hence, if we consider the end of inflation with $N_{\rm end } < 20$, then the $B_3$ term can be neglected in most cases of interest. Under this condition the solution of \eqref{eom_sigma_general} reads 
\begin{align}\label{sol_sigma_3}
\sigma_{3\pm}(N) =&\; C_{31} e^{-3N/2}\Gamma\left(1-2\frac{\nu_\sigma}{\delta_3}\right) 
J_{-\frac{2\nu_\sigma}{\delta_3}} \left(\frac{2}{\delta_3} \sqrt{A_{3\pm}e^{\delta_3 N}}\right) 
\nonumber\\
&+ C_{32} e^{-3N/2}\Gamma\left(1+2\frac{\nu_\sigma}{\delta_3}\right) 
J_{\frac{2\nu_\sigma}{\delta_3}} \left(\frac{2}{\delta_3}\sqrt{A_{3\pm}e^{\delta_3 N}}\right),
\end{align}
where $J_n(x)$ is Bessel function of the first kind and we have denoted $\nu_\sigma = \sqrt{\frac{9}{4} -M_\sigma^2}$. 
For each mass eigenstate, one can solve the coefficients $C_{31}$, $C_{32}$ by matching the solutions with boundary conditions $\sigma_{2\pm} = \sigma_{3\pm}$, $\frac{d\sigma_{2\pm}}{dN} =\frac{ d\sigma_{3\pm}}{dN}$ at $N = N_\ast$. The results are given by
\begin{align}
C_{31} =& \frac{\pi}{2\delta_3} \frac{e^{3N_\ast/2}}{\Gamma\left(1-\frac{2\nu_\sigma}{\delta_3}\right)} 
\left\{\left[-2 d\sigma_{2\ast} +(-3+2\nu_\sigma)\sigma_{2\ast}\right] 
J_{\frac{2\nu_\sigma}{\delta_3}} \left(\frac{2}{\delta_3}\sqrt{A_3 e^{\delta_{3}N_\ast}}\right) \right. \nonumber\\
&\left. -2\sigma_{2\ast}\sqrt{A_3 e^{\delta_{3}N_\ast}} 
J_{1+\frac{2\nu_\sigma}{\delta_3}} \left(\frac{2}{\delta_3}\sqrt{A_3 e^{\delta_{3}N_\ast}}\right)\right\} 
\csc\left(\frac{2\nu_\sigma}{\delta_3}\pi\right), 
\\\nonumber
C_{32} =& -\frac{e^{3N_\ast/2}}{4\nu_\sigma \delta_3^2} 
\left[\frac{1}{\delta_3}\sqrt{A_3 e^{\delta_{3}N_\ast}}\right]^{-\frac{2\nu_\sigma}{\delta_3}} 
\Gamma\left(-\frac{2\nu_\sigma}{\delta_3}\right) \times\\\nonumber
&\left\{2\delta_{3}\nu_\sigma \left[2\nu_\sigma\sigma_{2\ast} +(2d\sigma_{2\ast} +3\sigma_{2\ast})\right] 
{}_0\tilde{F}_1 \left[1-\frac{2\nu_\sigma}{\delta_3}; -\frac{A_3}{\delta_3^2} e^{\delta_3 N_\ast}\right] \right.\\
&\left.+ 4\nu_\sigma A_3 e^{\delta_3 N_\ast} \sigma_{2\ast} \;
{}_0\tilde{F}_1 \left[2-\frac{2\nu_\sigma}{\delta_3}; -\frac{A_3}{\delta_3^2} e^{\delta_3 N_\ast}\right]
\right\},
\end{align}
where ${}_0\tilde{F}_1[n;x] = \frac{{}_0F_1[n;x]}{\Gamma(n)}$ is the regularized confluent hypergeometric function, and $\sigma_{2\ast} \equiv \sigma_2(N_\ast)$, $d\sigma_{2\ast} \equiv \frac{d\sigma_2}{dN}\vert_{N = N_\ast}$ are given by \eqref{sol_sigma_2} and \eqref{sol_dsigma_2} respectively. It is important to note that we have suppressed the notation for each mass eigenstate in $C_{31}$, $C_{32}$, $A_3$, $\sigma_{2\ast}$ and $d\sigma_{2\ast}$.

We can estimate the (temporal) baryon asymmetry at the end of inflation by assuming instantaneous reheating soon after $N = N_{\rm end}$. The temperature $T_\ast = [\frac{30}{\pi^2 g_\ast}\rho_{r\ast}]^{1/4}$ is given by the energy density of radiation $\rho_{r\ast} \approx  3M_P^2 H_\ast^2$, where $g_\ast = 106.75$ is the number of relativistic degrees of freedom above $300$ GeV. The baryon asymmetry is then given by

\begin{align}\label{Y_B__analyitc_inflation_end}
Y_B(N_{\rm end}) = \frac{n_B(N_{\rm end})}{s_\ast} =\frac{H_\ast}{s_\ast} 
\left. \left(\sigma_{3+}\frac{d \sigma_{3-}}{d N} - \sigma_{3-}\frac{d \sigma_{3+}}{d N}\right)\right\vert_{N = N_{\rm end}},
\end{align} 
where $s_\ast = \frac{2\pi^2g_\ast T_\ast^3}{45}$ is the entropy production. According to the right panel of Figure~\ref{fig:M_3}, one can see that in the regime of $\delta_2 \ll -3$, the $A_3$ term in Eq.~\eqref{def_M_i} grows very fast in phase 3. If $M_3 \gg 1$ with sufficiently large $N_{\rm end}$, the AD field starts to oscillate rapidly, which washes away the generated baryon asymmetry.

\subsection{Final baryon asymmetry}\label{Sec_final_YB}
Having acquired the initial conditions of the AD field at the end of inflation, in this section we compute the final baryon asymmetry in the radiation domination epoch through reheating of the universe via inflaton decay. We consider that the decay of inflaton $\phi$, is dominated by a perturbative channel with the decay width $\Gamma_I$. As in the typical reheating scenario, inflation is terminated by a rapid oscillation of $\phi$ with an effective mass $\frac{\partial^2V(\phi)}{\partial\phi^2} \approx m_I^2 \gg H_\ast^2$ so that the inflaton density $\rho_I \sim a^{-3}$ decays as dust-like matter at the beginning of reheating when $H > \Gamma_I$. 

Denoting $t_{\rm end} = 0$ as the physical time at the end of inflation, we can write down an analytic solution in the limit of $H_\ast(t - t_{\rm end}) \gg 1$ as \cite{Wu:2021mwy,Wu:2020pej,Wu:2019ohx} 
\begin{align}\label{analytic_inflaton}
\phi(t) = \phi_{\rm max} a^{-3/2} \cos\left[m_I (t-t_{\rm end})\right] e^{-\Gamma_I (t-t_{\rm end})/2}, 
\end{align}
where $\phi_{\rm max} =\frac{ \Lambda_I^2}{m_I}$ is the maximal amplitude for $\phi$ at $t = t_{\rm end}$ and $\Lambda_I^4 = 3M_P^2H_\ast^2 = \rho_{I0}$ is the definition of the energy scale of inflation. It is easy to check that $t_r \equiv 1/\Gamma_I$ is the time scale at which the density of radiation $\rho_r$ starts to overcome $\rho_I$.

Using the background evolution $\square\phi = -\ddot{\phi} -3H\dot{\phi} \approx m_I^2\phi$ and $(\partial\phi)^2 = \dot{\phi}^2$ with $\phi(t)$ given by \eqref{analytic_inflaton}, we can solve the equations of motion for the mass eigenstates according to
\begin{align}
\ddot{\sigma}_{+} + 3H\dot{\sigma}_{+} + \left(m_\sigma^2 + \frac{c_1 + c_2}{\Lambda}m_I^2 \phi + \frac{c_3}{\Lambda^2}\dot{\phi}^2\right) \sigma_{+} &= 0, \\
\ddot{\sigma}_{-} + 3H\dot{\sigma}_{-} + \left(m_\sigma^2 + \frac{c_1 - c_2}{\Lambda}m_I^2 \phi + \frac{c_3}{\Lambda^2}\dot{\phi}^2\right) \sigma_{-} &= 0,
\end{align}
where initial conditions at the end of inflation ($N = N_{\rm end}$) are given by \eqref{sol_sigma_3}. The final baryon asymmetry at some time $t_f \gg t_r$ well inside the radiation dominated epoch reads
\begin{align}\label{Y_B_final}
Y_B(t_f) = \frac{n_B(t_f)}{s(t_f)} 
= \frac{45}{2\pi^2g_\ast T^3(t_f)} \left( \sigma_{+}\dot{\sigma}_{-} - \sigma_{-}\dot{\sigma}_{+}\right)_{t = t_f} .
\end{align}
Since initial conditions $\sigma_{\pm 0}$ and $\dot{\sigma}_{\pm 0}$ imported from \eqref{sol_sigma_3} depend on the USR parameters $\delta_2$, $N_\ast$ and $N_{\rm end}$, the final baryon asymmetry \eqref{Y_B_final} is therefore controlled by the inflationary scenario considered in Section~\ref{Sec_USR_inflation}.\footnote{Initial VEVs of the mass eigenstates can realise successful baryogenesis ($Y_B \sim 10^{-10}$) in the regime of $\sigma_{\pm 0} \ll H_\ast$. In contrast to the conventional scenario for AD baryogenesis from flat directions \cite{Dine:1995kz}, initial VEVs for the AD field are usually much larger than $H_\ast$.}

The value of $N_\ast$ for successful baryogenesis $Y_B \sim 10^{-10}$ is our main concern as it is the key parameter that determines the fraction of PBH density that comprises DM (see Section~\ref{Sec_PBH_DM}). To be more precise, we are interested in finding the threshold of $N_\ast$ (denoted as $N_{\ast c}$) from which the final baryon asymmetry is nearly unchanged, namely $Y_B(N_{\ast c}) \approx Y_B(N_\ast \gg N_{\ast c})$.
Our numerical tests for various choices of $\delta_2$ indicate that $N_{\ast c}$ is not sensitive to the AD mass for $m_\sigma \lesssim H_\ast$ but it is significantly affected by the cutoff scale $\Lambda$. In general, the smaller the cutoff $\Lambda$, the larger the value of $N_{\ast c}$ is obtained since the $A_2$ term in Eq.~\eqref{def_M_i} needs a longer time to be diluted. 

\begin{figure}
	\begin{center}
		\includegraphics[width=8cm]{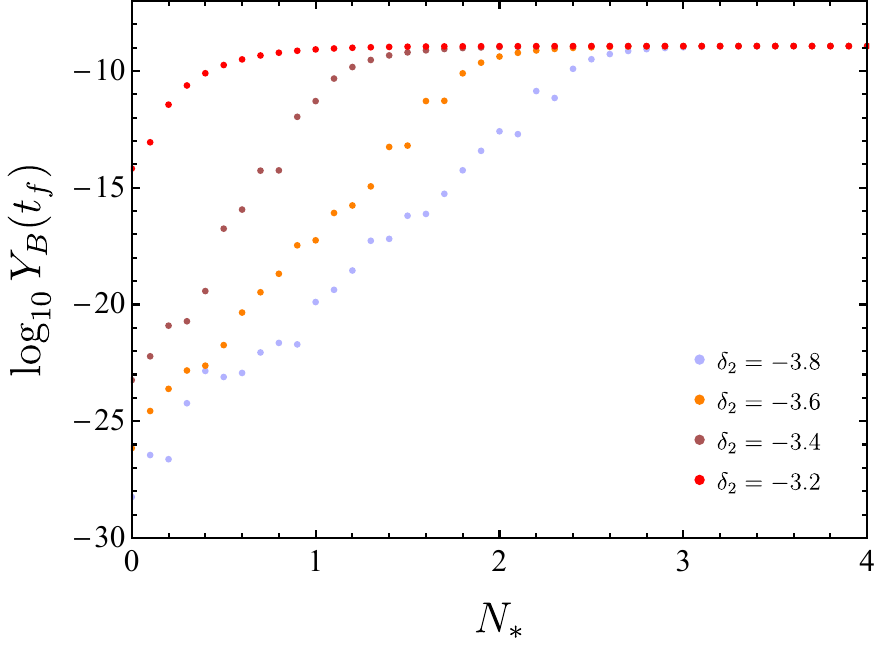} 
		%\hfill
		%\includegraphics[width=8cm]{Y_B_end.pdf} 
		%\par
	\end{center}
	\caption{ The final baryon asymmetry in radiation domination at $t=t_f$ with various choices of $\delta_2$ and $\delta_3 = -\delta_2 -3$, where $\frac{m_\sigma}{H_\ast} = 0.1$, $\frac{\Lambda}{M_P} = 1$ and $N_{\rm end} = 18$ is used.  \label{fig:Y_B_final}}
\end{figure}

In the rest of this work, we shall focus on the highest cutoff $\Lambda = M_P$ since this gives the lowest $N_{\ast c}$ possible among all the viable scenarios so that the parameter space for the constant $Y_B$ is maximized. As shown in Figure~\ref{fig:Y_B_final}, the constant $Y_B$ plateau is recovered in the USR limit when $\delta_2\to -3$, which are the cases investigated in Refs.~\cite{Wu:2021mwy,Wu:2021gtd}. With the generalization of $\delta_2$ away from $-3$, the constant $Y_B$ plateau is narrowed down towards large $N_\ast$. This result has important consequences for the ``cosmic coincidence'' as will be addressed in the next Section.

\section{The cosmic coincidence of dark matter and baryons}\label{Sec_CC}
CMB observations reported that the cold dark matter (CDM) density in the standard $\Lambda$CDM scenario today is $\Omega_{\rm CDM0} =0.265$ and the redshift at matter-radiation equality is $z_{\rm eq} = 3402$ \cite{Planck:2018vyg}. These findings indicate that $\Omega_{\rm CDMeq} = 0.42$ and $\Omega_{\rm Beq} = m_B n_{\rm Beq} = 0.08$, where $m_B = 0.938$ GeV is the averaged nucleon mass and $n_{\rm Beq} = \vert Y_B\vert s(t_{\rm eq})$ is the baryon number density. The specific ratio between the two species, namely
\begin{align}\label{ratio_CDM_B}
\frac{\Omega_{\rm CDMeq}}{ \Omega_{\rm Beq}} \approx 5,
\end{align}
seems to imply a non-trivial connection of their origin, since one would expect that the densities between CDM and baryons to differ by orders of magnitude if their genesis is completely uncorrelated in the early universe. Therefore the nearly $\mathcal{O}(1)$ ratio shown in Eq.~\eqref{ratio_CDM_B} is sometimes remarked referred to as the ``cosmic coincidence problem'' and in this section we address a possible scenario for the cosmic coincidence in which dark matter is composed by PBHs fostered from USR inflation.

\subsection{Primordial black hole dark matter}\label{Sec_PBH_DM} 

Since the inflaton decays during reheating, the enhanced curvature perturbations are inherited by the radiation density perturbations. Therefore, shortly after reentry into the horizon in radiation domination, PBHs are formed at the high variance peaks of the density fluctuations. This is because the overdense regions will cease expanding some time after they enter the particle horizon and collapse against the pressure if they exceed the the Jeans mass. 

We shall focus on the viable window for PBHs to comprise all DM in the asteroid-mass range $M_{\textrm{PBH}}\simeq 10^{-16}\hbox{-} 10^{-12} M_\odot$ which indicates that the pivot scale $k_0/k_{\textrm{eq}}\simeq(M_{\textrm{eq}}/M_{\textrm{PBH}})^{1/2}$ for the initiation of a USR phase in the range $k_0\simeq 3\times 10^{12}\hbox{-}10^{14}\,\textrm{Mpc}^{-1}$, where
$k_{\textrm{eq}}$ and $M_{\textrm{eq}}$ denote the horizon wavenumber and the horizon mass at matter--radiation equality respectively. The comoving horizon length is defined $R=1/(aH)$
and the horizon mass is given by $M_H = \frac{4\pi\rho}{ 3(H)^3}$. We will define $R$ as a function of $M_H$ below.

Given a specific primordial scalar power spectrum $\mathcal{P}_\zeta(k)$, we can use standard techniques to calculate the PBH mass function $f(M)$ which characterises the fraction of PBHs constituting DM today. Here, we shall focus on scales of the density field that reenter the Hubble radius during the radiation dominated epoch. At matter-radiation equality, the PBH-to-DM density ratio is given by
\begin{align}
\label{eq:fPBH}
f_\textrm{PBH}=\frac{\Omega_{\textrm{PBHeq}}}{\Omega_{\textrm{DMeq}}}=\frac{1}{\Omega_{\textrm{DMeq}}}\int f(M_H)d\log M_H,
\end{align}
where $f(M_H)=\frac{\beta(M_H)}{\Omega_\textrm{DM}}\left(\frac{M_{\textrm{eq}}}{M_H}\right)^{1/2}$ shows the PBH mass function as a function of the horizon mass at matter-radiation equality and $\beta(M_H)$ is the fraction of PBH density at the given $M_H$.\footnote{Strictly speaking, the PBH mass $M_{\rm PBH}$ can be much smaller than the horizon mass $M_H$ at the formation epoch due to the effect of critical collapse \cite{Niemeyer:1997mt,Yokoyama:1998xd}. Therefore the density fraction $\beta = \beta(M_{\rm PBH}, M_H)$ in general has a distribution in $M_{\rm PBH}$ at a given $M_H$. However, if we are only interested in the value of $N_\ast$ for a fixed ratio for example, $f_{\rm PBH} = 1$, then the difference led by critical collapse is negligible \cite{Wu:2021gtd}. 
	%even if one uses the most broadest spectrum very close to the USR limit ($\delta_2 \rightarrow −3$), the difference in $M_{\rm PBH}$ in terms of $N_\ast$ is negligible \cite{Wu:2021gtd}. 
	Thus, for USR inflation, we find that $M_{\textrm{PBH}} \approx M_H$ can be a good approximation for resolving the PBH abundance and this relation simplifies the density fraction as $\beta = \beta(M_H)$. }
Note that PBHs behave as dust-like matter and thus $a_{\textrm{eq}}/a \sim (M_{\rm eq}/M_H)^{1/2}$ describes the relative growth of $\beta(M_H)$ during radiation domination.
 The scale dependent horizon mass $M_H$ and the co--moving scale $R$ can be written in terms of $M_H$ as
\begin{align}
\label{eq:MHandRH}
    M_H &= M_{\textrm{eq}}\left(\frac{k_{\textrm{eq}}}{k}\right)^2\left(\frac{g_{\textrm{eq}}}{g_{*}}\right)^{1/3},& R(M_H) &= \frac{1}{k_{\textrm{eq}}}\left(\frac{M_H}{M_{\textrm{eq}}}\right)^{1/2}\left(\frac{g_{*}}{g_{\textrm{eq}}}\right)^{1/6},
\end{align}
where the horizon mass is given by $M_{\textrm{eq}}=2.94\times 10^{17}M_{\odot}$ and the number of relativistic degrees of freedom at matter--radiation equality is given by $g_{\textrm{eq}}=3$, it also follows from Eq.~\eqref{eq:MHandRH} that $R(M_{\textrm{eq}})\approx 1.57\times 10^{40}\,\textrm{GeV}^{-1}$. We use $k_{\textrm{eq}}=0.01\, \textrm{Mpc}^{-1}$ and $g_{*}=106.75$ since the horizon mass satisfies $M_H<1.5\times 10^{-7} M_{\odot}$ where the temperature of the universe exceeds $300$ GeV.

If we now consider the simplest fiducial Press-Schechter statistic \cite{Carr:1975qj} for Gaussian fluctuations to be distributed with a dispersion $\sigma_0$, then the rare density peaks that exceed a critical value $\delta_c$, are responsible for production of PBHs. The fraction of energy density that collapse into BHs at the mass scale $M_H$ during radiation domination reads
\begin{align}\label{def_beta_PS}
\beta(M_H)=2\int_{\nu_c}d\nu\frac{e^{-\nu^2/2}}{\sqrt{2\pi}}  
= \textrm{erfc}\left(\frac{\nu_c}{\sqrt{2}}\right)
= \textrm{erfc}\left(\frac{\delta_c}{\sqrt{2}\sigma_0}\right),
\end{align}
where $\textrm{erfc}(z)$ is the complimentary error function, $\nu = \delta_r/\sigma_0$ is the peak value of radiation fluctuations, and we take the critical value $\delta_c =0.45$ from \cite{Harada:2013epa}. Based on the linear relation $P_{\delta_r} = (\frac{4}{9})^2(kR)^4P_\zeta$ in our fiducial estimation, the variance $\sigma_0^2(R)$ for a given power spectrum $\mathcal{P}_\zeta$ is computed by
%\textcolor{red}{
\begin{align}\label{def_variance}
    \sigma_0^2(R)&=\int_{0}^{\infty} \frac{dq}{q} \, \left(\frac{4}{9}\right)^2\left[\left(qR\right)^4 P_\zeta(q) W^2(qR) T^2(qR)\right],
\end{align}
%}
% \begin{align}
%     \sigma^2(R)&=\frac{16}{81}\int_{0}^{\infty} d\log k \, \left[k^4 R^4 \mathcal{P}_R(k) W^2(k,R)\right]
% \end{align}
where the window function $W(qR)$ and the transfer function $T(qR)$ contained in the kernel of integral are given by
\begin{align}\label{def_window_transfer}
  W(qR)=e^{-\frac{q^2 R^2}{2}}, \qquad 
  T(qR) = \frac{9\sqrt{3}}{(qR)^3}\left(\sin\frac{q R}{\sqrt{3}}-\frac{q R}{\sqrt{3}}\cos \frac{q R}{\sqrt{3}}\right),
\end{align}
respectively.
Here we choose the volume-normalised Gaussian window function with smoothing radius $R = 1/(aH)$ led by the comoving horizon at conformal time $\eta$. Therefore one can refer to $R$ or $M_H$ as the time parameter in this calculation. Note that $T(qR) \rightarrow 1$ as $qR \rightarrow 0$ where the $q$ mode perturbation is well outside the horizon scale.
%For a given wave number $k$ corresponding to horizon reentry during radiation domination, we have $=\eta^{-1}=aH=\mathcal{H} $ and $R=\mathcal{H}^{-1}=k^{-1}$. Hence from Ref.~\cite{Cai:2019elf}, we get

%\textcolor{red}{
%\begin{align}
 %   \sigma^2(R)&=\frac{16}{81}\int_{0}^{\infty} \frac{dq}{q} \, \Bigg\{\left(q R\right)^4 \mathcal{P}_R(q) W^2(q,R) \left[\frac{9\sqrt{3}}{(qR)^3}\left(\sin\frac{q R}{\sqrt{3}}-\frac{q R}{\sqrt{3}}\cos \frac{q R}{\sqrt{3}}\right)\right]^2\Bigg\}
%\end{align}
%}

%This analytical function is not of a purely Gaussian form and has a smoothing radius  and prevents divergence in the limit $k\rightarrow 0$. 

%%%%%%%%%%%%%%%%%%%%%%%%%%%%%%%%%%%%%
\begin{figure}[!h]
    \centering
     \includegraphics[width=0.48\textwidth]{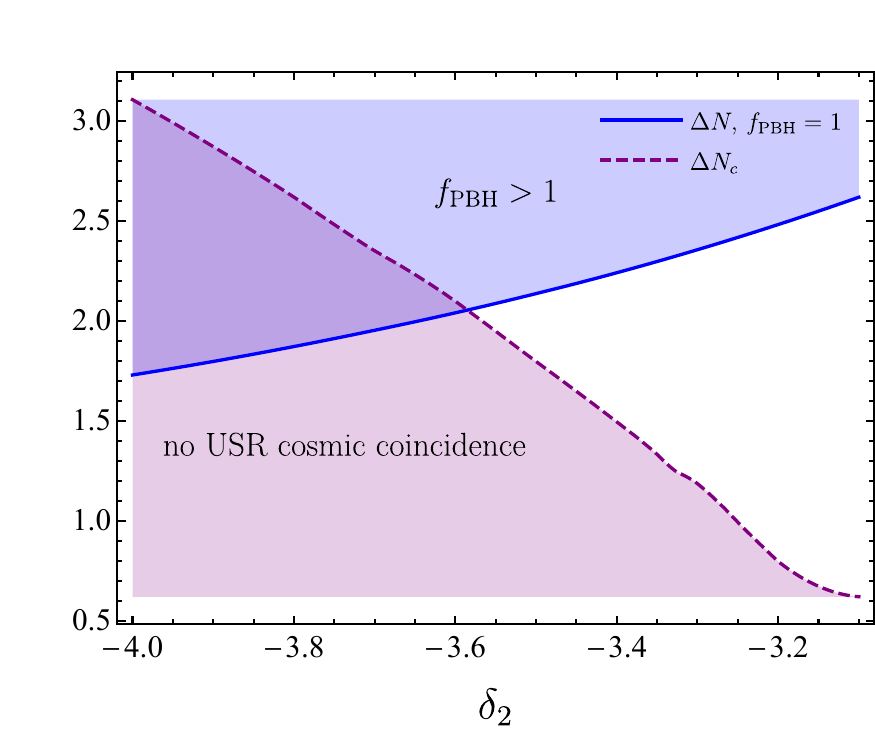}
         \includegraphics[width=0.505\textwidth]{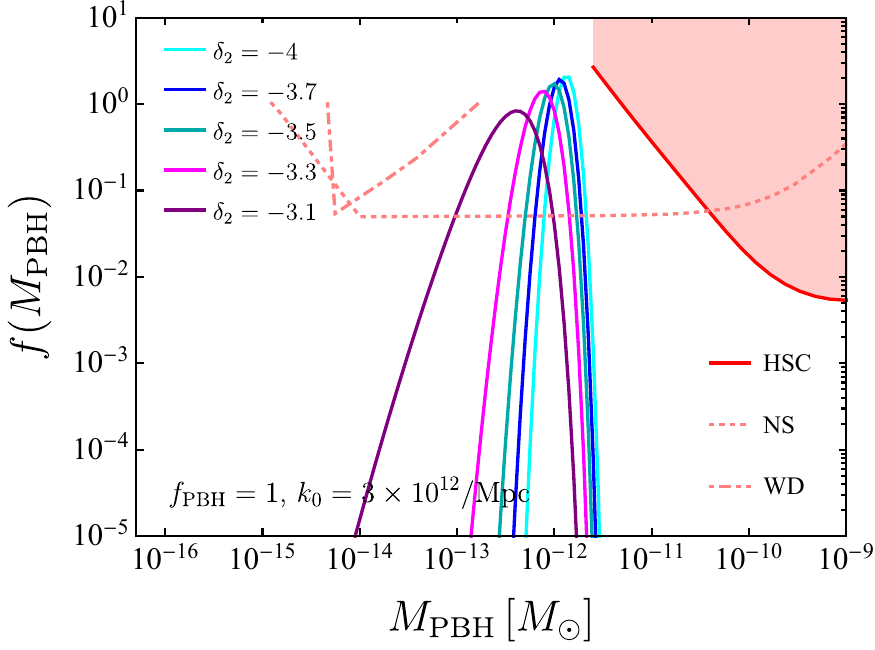}\\
    \caption{The ultra-slow-roll inflation duration $\Delta N$ (solid line) required for primordial black holes to saturate the total dark matter density and $\Delta N_c$ (dashed) required for the cosmic coincidence is shown as a function of the ultra-slow-roll rate $\delta_2$ (left panel). The blue and purple shaded regions correspond to ``overabundant primordial black hole dark matter'' and ``no clear indication to the cosmic coincidence problem from the USR triggered baryogenesis'' respectively. The primordial black hole mass functions for various rates are shown  (right panel), we consider a broken power-law spectrum with a pivot scale of $k_0=3\times 10^{12}\, \textrm{Mpc}^{-1}$ rates of $\delta_2=-[4,3.7,3.5,3.3,3.1] $. The constraints coming from neutron stars (NSs) and white dwarfs (WDs) are shown in dotted and dot-dashed respectively while the solid line with the upper shaded region corresponds to HSC M31 microlensing limits which are obtained from the PlotBounds \cite{bradley_j_kavanagh_2019_3538999} package. }
    \label{fig:PBHmassfunctions}
\end{figure}
%%%%%%%%%%%%%%%%%%%%%%%%%%%%%%%%%%%%%

Let us now apply the generalised USR scenario \eqref{USR_template} to \eqref{def_variance} to obtain the mass function $f(M)$. Note that $P_\zeta$ is only controlled by $\delta$ and $\Delta N$ and therefore $f(M)$ or $f_{\rm PBH}$ is not sensitive to $N_{\rm end}$.
In Figure~{\ref{fig:PBHmassfunctions}}, we show the USR duration $\Delta N = N_\ast$ as function of the USR rate $\delta_2$ while ensuring that PBHs produced through USR inflation in Eq.~\eqref{eq:fPBH} saturates the DM density limit ($f_{\rm PBH} = 1$). The explicit PBH mass functions in Figure~{\ref{fig:PBHmassfunctions}} are shown for a selection of $\delta_2$ values. We have also shown constraints for the large mass limit of the ultralight asteroid-mass window from neutron star disruption \cite{Capela:2013yf} and the white dwarf explosions triggering supernovae \cite{Graham:2015apa} as dashed and dot--dashed curves. These stellar disruption limits are not currently considered robust since having being discredited by Ref.~\cite{Montero-Camacho:2019jte}. Improving experimental limits in this mass range presents an excellent way to test the mass spectrum of PBHs produced by the cosmic coincidence. The more robust microlensing constraints for sub--planetary--mass compact objects, including PBHs, coming from Subaru HSC observations of M31 \cite{Smyth:2019whb} are shown as the red shaded region in the right panel. This excludes PBH DM saturation above $10^{-12}M_\odot$ but loses strength at lower mass. The selection of the pivot scale $k_0$, the USR rates $\delta_2$, and durations $\Delta N$, largely evade the best available constraints. 

% We can then make use of the above relation between R and M and the expression for β(M) to finally arrive at
% fPBH (M) utilizing eq. (3.1). It is well known that the threshold of the density contrast δc is
% a crucial parameter since fPBH is exponentially sensitive to it. The value of δc is expected
% to lie in the range 0.3–0.65 (see refs. [74–77], see however the recent discussion [78]). For
% the purposes of illustration, we shall work with δc = 1/3 and 0.5. We should clarify that
% the exact value of this parameter does not affect the primary conclusions we draw about the
% mechanism of generating PBHs from squeezed initial states.

\subsection{The cosmic coincidence}
\label{sec:cosmiccoincidence}
In the regime close to the exact USR inflation scenario ($-3.2 < \delta_2 < -3$), the required duration for PBHs to account for all DM ($\Delta N > 2$, see the left panel of Figure~{\ref{fig:PBHmassfunctions}}) lies well within the constant $Y_B$ plateau, as shown in Figure~\ref{fig:Y_B_final}. Given that a $\mathcal{O}(0.2)$ variation in $\Delta N$ can cause around $10^{50}$ difference in $f_{\rm PBH}$ \cite{Wu:2021gtd}, the uncertainty of $\Delta N$ (for $f_{\rm PBH} = 1$) due to various effects in the statistics of PBH abundance is at most of $\mathcal{O}(10^{-1})$ \cite{Wu:2021mwy}. This is the reason why the coincidence ratio $\frac{\Omega_{\textrm{DMeq}}}{ \Omega_{\rm Beq}} \approx 5$ is rather guaranteed in this scenario despite the fact that we are using the simplest statistical method shown in Eq.~\eqref{def_beta_PS}.  

However, with the extension of the scenario to the regime of $\delta_2 < -3.2$, the constant $Y_B$ plateau starts to collapse towards the large $\Delta N$ limit (see Figure~\ref{fig:Y_B_final}), whereas the required $\Delta N$ for $f_{\rm PBH} = 1$ is smaller since the enhancement for $A_{\rm PBH}$ given in Eq.~\eqref{USR_template} is more efficient. As a result, one expects that the parameter space for PBH DM to drop out of the $Y_B$ plateau with a sufficiently small $\delta_2$, wherein the scenario loses its indication of the cosmic coincidence as the correct baryon asymmetry also relies on tuning of other parameters.  

To specify the lower bound of $\delta_2$ for preserving the predictability of the cosmic coincidence, let us quantify the plateau threshold $\Delta N_c = N_{\ast c}$ by defining the edge of the constant plateau of the final baryon asymmetry as $Y_B(N_{\ast c}) \equiv Y_B(N_\ast \rightarrow \infty)/2$. For a numerical demonstration we take the asymptotic $Y_B$ at $N_\ast = 4$. The results based on this definition are given in the left panel of Figure~{\ref{fig:PBHmassfunctions}}, where the intersection is around $\delta_2 = -3.6$. This implies that the required $\Delta N$ for $f_{\rm PBH} = 1$ is located inside the constant $Y_B$ plateau when $\delta_2 > -3.6$. Considering the uncertainties in the statistics of PBH abundance, we therefore impose conservative bounds for the USR rate as
\begin{align}\label{CC_condition}
-3.5 < \delta_2 < -3.1  \qquad (\textrm{for the cosmic coincidence}),
\end{align}
where the conservative upper bound ensures that the effective mass of inflaton during the USR phase (phase 2) is at least of order $H_\ast$ so that the effect of quantum diffusion can be largely suppressed \cite{Ng:2021hll}.
%extended mass function does not touch the evaporation constraints in the small PBH mass limit. 

\subsection{Implications for the gravitational wave background}
In this section we study the relevant GWs generated in the PBH dark matter scenario and we also investigate the implications from the cosmic coincidence to the resulting GW spectrum for current and future observations.

\subsubsection{Induced gravitational wave production}
\label{sec:IGWproduction}
Here, we are interested in calculating the GW spectral density as measured in the present universe. For a given mode in Fourier space, the frequency of GWs today is given by
\begin{align}
    f=\frac{k}{2\pi}=1.55\times 10^{-15} k \, \textrm{Mpc} \, \textrm{Hz}.
\end{align}
With the scale crossing the Hubble radius at matter-radiation equality being $k_{\textrm{eq}}=1.3\times 10^{-2}\textrm{Mpc}^{-1}$
, all modes with frequencies $f \gtrsim 10^{-17}$ Hz have re-entered the horizon during radiation domination, 
assuming the standard thermal history between the end of inflation and the radiation-dominated era considered in Section~\ref{Sec_final_YB}.
%unless a non-standard thermal history is considered between the end of inflation and the radiation-dominated era. 
For the broken power-law type spectrum adopted in Eq.~\eqref{USR_template}, the peak frequency $f_{\textrm{peak}}$ of the GW spectrum is determined by the peak scale of the curvature perturbation spectrum $\mathcal{P_\zeta}$ (namely at pivot scale $k_0$), which is related to the horizon mass corresponding to the pivot scale as \cite{Cai:2019elf}
 \begin{align}
     f_{\textrm{peak}} \sim 6.7\times10^{-9} \left(\frac{M_{H}(k_0)}{M_{\odot}}\right)^{-1/2} \textrm{Hz}.
 \end{align}
Note that the central peak in the PBH mass function depends on the choice of $\delta_2$ as one can see from the right panel of Figure~\ref{fig:PBHmassfunctions}. 
 
The induced GWs sourced by the scalar-mode perturbation (on cosmological scales) at second order consist a stochastic background, and it is usual to describe the induced GW spectrum by energy density per logarithmic frequency interval normalized by the critical density. The dimensionless spectral density associated with the induced GWs, $\Omega_{\textrm{IGW}}(k, \eta)$, evaluated at late enough times when the modes are contained within the Hubble radius during the radiation dominated epoch, is given by
\begin{align}
\label{eq:GWspectrum}
\Omega_{\textrm{IGW}}(k,\eta)&=\frac{1}{24}\left(\frac{k}{aH}\right)^2 \overline{P_h(k,\eta)},
\end{align}
where the conformal time is defined $\eta=(aH)^{-1}$ at horizon reentry in the radiation dominated era and the two respective polarisation modes of GWs have been summed over. $P_h$ is the power spectrum of the induced tensor-mode perturbation sourced by linear scalar-mode perturbations at second order \eqref{def_P_h}, which can be solved via the Green's function method \cite{Ananda:2006af,Baumann:2007zm} as
\begin{align}\label{sol_h_Green}
h_\lambda(\vec{k},\eta) = 4 \int^{\eta} d\eta_1 G_{\vec{k}}(\eta;\eta_1) \frac{a(\eta_1)}{a(\eta)} S_\lambda (\vec{k},\eta_1),
\end{align}
where $\lambda = +, \times$ are the two polarisations, $G_{\vec{k}}(\eta;\eta_1) =\frac{1}{k}\sin(k(\eta -\eta_1))$ is the Green's function in radiation domination and $S_\lambda$ is the source term with detailed information provided in Appendix~\ref{Appd_tensor_spectrum}.
The overline in Eq.~\eqref{eq:GWspectrum} denotes average over a few wavelengths for time oscillations led by the Green's function and the transfer function $T(k\eta)$ given in Eq.~\eqref{def_window_transfer}.
Note that the GW density measured in today's universe, $\Omega_{\textrm{IGW0}} h^2$, is related to \eqref{eq:GWspectrum} in radiation domination like
\begin{align}
\Omega_{\textrm{IGW0}} h^2 &= \frac{h^2}{3M_P^2 H_0^2} \frac{d\rho_{\rm IGW}}{d \ln k}\nonumber \\
&= \Omega_{r0}h^2\frac{\rho_r}{\rho_{r0}}\left(\frac{a}{a_0}\right)^4 \Omega_{\textrm{IGW}} \nonumber \\
&\approx 1.62 \times 10^{-5}  \Omega_{\textrm{IGW}}, 
\end{align}
where $\Omega_{r0} h^2= 4.18 \times 10^{-5}$ is the density fraction of radiation today \cite{Planck:2018vyg} and the factor $\frac{\rho_r(\eta)}{\rho_{r0}}\left(\frac{a(\eta)}{a_0}\right)^4 = \frac{g_\ast}{g_0}\left(\frac{g_\ast}{g_0}\right)^{-4/3} \approx 0.4$ \cite{Bartolo:2018evs,Bartolo:2018rku} is valid for $\eta \ll \eta_{\rm eq}$ where the temperature of the universe is much higher than 300 GeV. This is correct in our case since the epoch $\eta_{k_0} = -1/k_0$ at which the pivot scale reenters the horizon for the PBH masses of interest ($k_0 = 10^{12}-10^{14}$ Mpc$^{-1}$) corresponds to the temperature $T \simeq 10^6 - 10^7$ GeV. 

% Using the relation between scalar perturbations $\Phi$ and curvature perturbations $\mathcal{R}$ in the radiation-dominated era, $\Phi = -(2/3)\mathcal{R}$, we can calculate the power spectrum induced by the curvature perturbation during radiation-dominated era as \cite{Kohri:2018awv}

%%%%%%%%%%%%%%%%%%%%%%%%%%%%%%%%%%%%%
\begin{figure}[!h]
    \centering
     \includegraphics[width=0.75\textwidth]{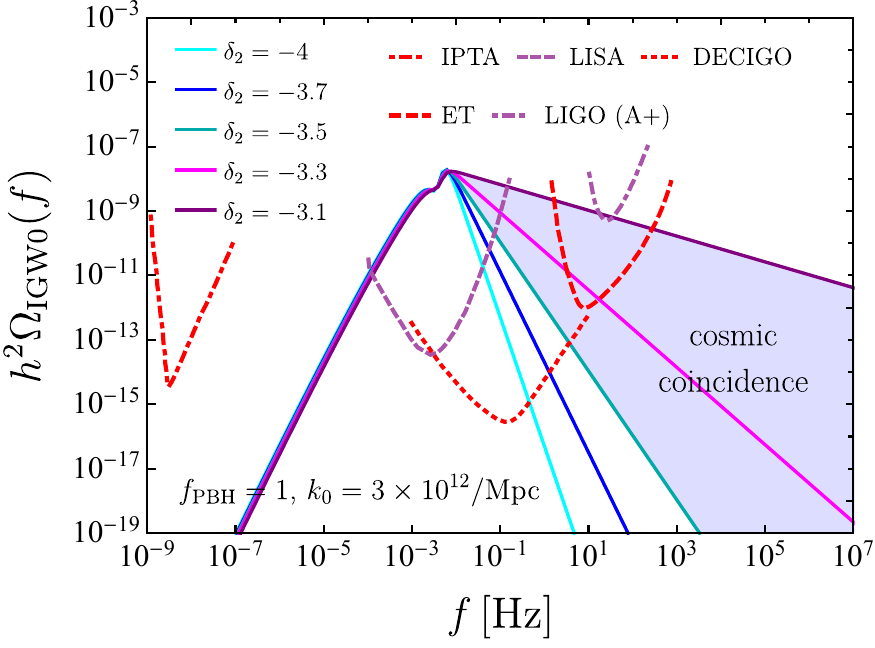}
    \caption{The present day dimensionless induced gravitational wave spectrum $h^2\Omega_{\textrm{IGW0}}$ is shown as a function of frequency $f$ in Hz assuming radiation domination for ultra--slow--roll rates of $\delta_2=-[4,3.7,3.5,3.3,3.1] $ are shown (solid lines). The shaded region between $\delta_2=-3.1$ and $\delta_2=-3.5$ corresponds to the cosmic coincidence region. The theoretical predictions are overlaid with relevant experimental constraints coming from IPTA (red dot-dashed), LISA (purple dashed), LIGO (A+) (purple dot-dashed), ET (red dashed) and DECIGO (red dotted). The experimental bounds are derived from the power-law integrated sensitivity curves \cite{Thrane:2013oya} in terms of the gravitational wave spectral density for some gravitational wave detectors universe at the time of gravitational wave generation. In each case we consider a broken power-law power spectrum with pivot scale $k_0=3\times 10^{12}\, \textrm{Mpc}^{-1}$.}
    \label{fig:IGWspectrum}
\end{figure}

Using \eqref{USR_template} as an input for the primordial value of the source term, we show the result of \eqref{eq:GWspectrum} in Figure~\ref{fig:IGWspectrum} for the induced gravitational wave spectrum at present day for the same USR rate $\delta_2$ and duration $\Delta N$ range considered in Section~\ref{Sec_PBH_DM}. We emphasise once again, that the PBHs produced during inflation ensures that the DM density of the universe is completely explained by PBHs. For the chosen pivot scale $k_0 = 3 \times 10^{12}$ Mpc$^{-1}$, we find the peak frequency to be around $f_{\textrm{peak}}\simeq 7\times 10^{-3}$ Hz with a peak amplitude around $\simeq 10^{-8}$.\footnote{For an input inflationary spectrum parametrised by a lognormal template with the spectral amplitude $A_\zeta$ and width $\Delta$, we find the ratio $\frac{\Omega_{\textrm{IGW}}}{A_\zeta^2} < 0.01$ for the entire range of the induced GW spectrum, except for the resonance peak in the limit of $\Delta \rightarrow 0$ where the lognormal template reproduces the delta-function spectrum \cite{Pi:2020otn}. This is the reason why the peak amplitude of $\Omega_{\textrm{IGW}}^{\rm peak}\approx A_{\rm PBH}^2$ led by broken power-law templates considered in this work is higher than induced GW spectra provided in Refs.~\cite{Domenech:2021ztg,Bartolo:2018evs,Bartolo:2018rku,Cai:2018dig} for PBHs as all DM.} 
The peak in $\Omega_{\textrm{IGW}}$ near $k \sim 2k_0/\sqrt{3}$ is due to the resonance (of the Green's function) of the tensor perturbation with both of the scalar perturbation sources \cite{Domenech:2021ztg}, where $1/\sqrt{3} = c_s$ is the sound speed of radiation and the factor two captures the second-order effect.

The induced GW spectral shape from scalar perturbations with a broken power-law spectrum has been well studied, which is indeed the case for our input in Eq.~\eqref{USR_template}. In what follows we summarise the model-(in)dependent results with respect to the USR parameters.

\bigbreak
\noindent
\textbf{The infrared (IR) tail:}
In the regime of $k \ll k_0$, the only contribution from our input template is $P_\zeta = A_{\rm PBH}(k/k_0)^{n_{\rm IR}}$ with $n_{\rm IR} = 4$ in the range of $k_{\rm min} < k < k_0$. This is the soft limit for the induced tensor perturbation where the main contribution is from scalar modes with $k \approx k_0$ that have been well inside the horizon. 
One can check that the scaling dependence in the integrand of \eqref{eq:P_h_uv_variable} only comes from the transfer function $T(k\eta)$ given in Eq.~\eqref{def_window_transfer}, as the scalar perturbations well inside the horizon behave like plane waves. In our case, with $n_{\rm IR} = 4$, the convolution of internal momentum of the scalar sources is dominated at the peak scale $k_0$ \cite{Atal:2021jyo}, where it is found that
\begin{align}
\Omega_{\textrm{IGW}}(k\ll k_0) \sim A_{\rm PBH}^2 \left( \frac{k}{k_0}\right)^3 \ln^2 \left(\frac{k}{k_0}\right).
\end{align}
The $\Omega_{\textrm{IGW}} \sim \left(\frac{k}{k_0}\right)^3 \sim f^3$ behaviour manifests the stochastic nature of the signals as if the GW background is a kind of random white noise \cite{Cai:2019cdl}. The logarithmic running $\sim \ln^2(k/k_0)$ is due to the integration over the oscillatory part of the transfer function \eqref{def_window_transfer}, which is a characteristic feature of the induced GW spectrum generated during the radiation dominated universe \cite{Yuan:2019wwo}. Note that both the $k^3$ scaling and the logarithmic running are model independent features for the input spectrum $P_\zeta$ so that the resulting $\Omega_{\textrm{IGW}}$ exhibits little difference among various choices of $\delta_2$.

\bigbreak
\noindent
\textbf{The ultraviolet (UV) tail:}
In the regime of $k \gg k_0$, our input template reads $P_\zeta = A_{\rm PBH}\left(\frac{k}{k_0}\right)^{-n_{\rm UV}}$, where the USR parameter enters the spectral index as $n_{\rm UV} = -2\delta_2 -6$. This regime corresponds to the generation of GWs right after horizon reentry for the scalar modes so that there is not enough time for the Green's function or the transfer function to play an important role. While the results depend on the choice of $n_{\rm UV}$ in general \cite{Domenech:2019quo, Domenech:2020kqm,Atal:2021jyo}, we focus on cases with $n_{\rm UV} < 4$ (due to the cosmic coincidence condition in Eq.~\eqref{CC_condition} for the USR rate $\delta_2$), in which the GW spectrum behaves as if it directly extracts the momentum dependence from the input power spectrum
\begin{align}
\Omega_{\textrm{IGW}}(k \gg k_0) \sim A_{\rm PBH}^2 \left( \frac{k}{k_0}\right)^{-2n_{\rm UV}(\delta_2)}.
\end{align} 
The cosmic coincidence conservative shaded parameter region between $-3.5\leq\delta_2\leq-3.1$ in Figure~\ref{fig:IGWspectrum} corresponds to the condition \eqref{CC_condition} described in Section~\ref{sec:cosmiccoincidence}. We remark that baryogenesis triggered by USR inflation is still available even if the UV tail of the induced GW spectrum is found to be outside the shaded region. It is only that both the observed baryon asymmetry and the PBH DM rely on fine-tuning of parameters so that the scenario discussed in Section~\ref{Sec_B_from_USR} has no clear resolution of the cosmic coincidence. 

%In each case, we note that the infrared tail and peaks of the various USR scenarios shown show little $\delta_2$ dependence while the ultraviolet tail shows strong $\delta_2$ dependence. The higher $\delta_2$ is, the higher the amplitude of the ultraviolet tail of the gravitational wave spectrum. \textcolor{blue}{The cosmic coincidence conservative shaded parameter region between $-3.5\leq\delta_2\leq-3.1$ corresponds to the cosmic coincidence described in Section~\ref{sec:cosmiccoincidence}.}

\bigbreak
\noindent
\textbf{Observational implications:}
The predictions for the USR inflation model considered yields compelling GW phenomenology in the observational window of the aforementioned experiments. The relevant GW bounds include those determined by studying the time of arrival from many pulsars in space. This includes data from the pulsar timing array (PTA) \cite{1979ApJ...234.1100D} which is comprised of three constituent projects, the European Pulsar Timing Array (EPTA) \cite{Desvignes:2016yex}, the Parkes Pulsar Timing Array
(PPTA) \cite{Hobbs:2013aka}, and the North American Observatory for Gravitational Waves (NANOGrav) \cite{McLaughlin:2013ira}, while the International Pulsar Timing Array (IPTA) \cite{IPTA2016} constraint comes from a combination of all three and covers the frequency band $10^{-9}$--$10^{-7}$ Hz. At higher frequency, we have a band that would be observable with the advanced Laser Interferometer Gravitational Wave Observatory (aLIGO) \cite{Thrane:2013oya} and Einstein Telescope (ET) \cite{Maggiore:2019uih}, which would be sensitive to the range $10$--$10^3$ Hz. In between IPTA and aLIGO/ET we expect LISA \cite{amaroseoane2017laser,Barausse:2020rsu} and DECIGO \cite{Yagi:2011wg,Kawamura:2020pcg} to be applicable.
In Figure~\ref{fig:IGWspectrum}, we display the recently derived future sensitivity curves for ET and LISA using the latest experimental design specification computed in Ref.~\cite{Bavera:2021wmw}. These are obtained by including the stochastic GW background, in contrast to model selection analyses that have discriminating power only up to the GW detector horizons. The use of the stochastic GW background to determine sensitivity curves accounts for the redshift integration of all GW signals in the universe and therefore presents a more appropriate projection for comparison with model predictions.

The current bounds on the stochastic GW background from joint LIGO and Virgo O3 data excludes $\Omega_{\textrm{IGW}} > 5.8\times 10^{-9}$ in the band of $20$-$76.6$ Hz at $2\sigma$ level \cite{KAGRA:2021kbb}. Note that this is the result for the flat GW spectrum ($\alpha = 0$), from the power-law method
\begin{align}
    \Omega_{\rm GW}(f) = \Omega_{\rm ref}\left(\frac{f}{f_{\rm ref}}\right)^{\alpha},
\end{align}
where $f_{\rm ref} =25$ Hz and $\alpha = 2/3$ describes the stochastic background from compact binary mergers \cite{LIGOScientific:2019vic,KAGRA:2021kbb}. The condition for the ``cosmic coincidence'' in Eq.~\eqref{CC_condition} for the presented scenario translates to $-2 < \alpha < -0.2$, where the posterior strength is $\Omega_{\rm ref} < 10^{-8}$ in this region by using the log-uniform prior. 
%The LIGO O3 constraint means that the entire UV tails in the range of \eqref{CC_condition} would have been ruled out for the USR spectrum with a pivot scale larger than $k_0 \approx 5\times 10^{13}$ Mpc$^{-1}$ (that is for PBH DM with masses smaller than $3.6\times 10^{-15} M_{\odot}$). For a pivot scale as small as $k_0 = 3\times 10^{12}$ Mpc$^{-1}$ (which corresponds to a peak PBH mass around $10^{-12} M_{\odot}$), O3 constraints on the UV tails shown in Figure~\ref{fig:IGWspectrum} indicates that the USR rate in the range of $-3.5 < \delta_2 < -3.35$ is still valid for explaining the cosmic coincidence problem, while the projected design sensitivity of aLIGO (A+) could rule out the entire region down to $\delta_2 = -3.5$.
As shown in Figure~\ref{fig:IGWspectrum}, the UV tail of the induced GW spectrum which is realised as a consequence of PBH saturating DM (in the ultralight asteroid mass window) from USR inflation, could be tested by the projected design sensitivity of LIGO (A+), LISA, ET, DECIGO and TianQin \cite{Liang:2021bde}. A measurement of the spectral index $\alpha$ for the stochastic GW background in the range of $-2 < \alpha < 0$ would further support the cosmic coincidence triggered by USR baryogenesis. Non-detection of the stochastic GW background by LIGO (A+) could exclude down to $\delta_2 \simeq -3.2$ for $k_0 \gtrsim  10^{14}$ Mpc$^{-1}$, which corresponds to a peak PBH mass smaller than $10^{-15} M_{\odot}$. Note that LISA, TianQin and DECIGO would be able to test the IR tail and the peak position of $\Omega_{\textrm{IGW0}}$ for arbitrary $k_0$ chosen within the PBH mass window of interest $10^{-12}-10^{-16} M_{\odot}$.

%The current bounds on the stochastic GW background from combined LIGO O1 and O2 data excludes $\Omega_{\textrm{IGW}} > 6\times 10^{-8}$ in the band of $20$-$86$ Hz at $2\sigma$ level \cite{LIGOScientific:2019vic,Christensen:2018iqi}. This means that the entire UV tails in the range of the cosmic coincidence \eqref{CC_condition} would have been ruled out for the USR spectrum with a pivot scale larger than $k_0 \approx 10^{14}$ Mpc$^{-1}$. For a pivot scale as small as $k_0 = 3\times 10^{12}$ Mpc$^{-1}$, LIGO constraints on the UV tails shown in Figure~\ref{fig:IGWspectrum} indicates that the USR rate in the range of $-3.5 < \delta_2 < -3.3$ is still valid for explaining the cosmic coincidence problem, while the projected design sensitivity of aLIGO could rule out the entire region down to $\delta_2 = -3.5$.  

\subsubsection{Non-Gaussian corrections 
%	to the induced gravitational wave spectrum
}\label{Sec_non_G}

%\textcolor{blue}{
	In Section~\ref{sec:IGWproduction} above, we computed the GWs induced by a broken power-law primordial curvature power
spectrum. Previously, it has been argued that the spectral tilt of the USR spectrum $P_\zeta$ in the UV regime ($k\rightarrow \infty$) is related to the magnitude of the non-Gaussianity parameter $f_{\textrm{NL}}$ \cite{Atal:2018neu} and further argued that $f_{\textrm{NL}}$ may be inferred directly from measurement of the UV tail of the induced GW spectrum \cite{Atal:2021jyo}. Here, $f_{\rm NL}$ is defined according to the standard local ansatz 
\begin{align}
\zeta = \zeta_g + \frac{3}{5}f_{\rm NL} \zeta_g^2,
\end{align}
where $\zeta_g$ is the Gaussian curvature perturbation considered in Section~\ref{Sec_USR_inflation}.
%There are also some purely non--Gaussian contributions to the induced GWs that arise from the 4--point correlation function. 

Similar to the generation of secondary tensor perturbations, the second-order curvature perturbations can be generated by non-vanishing 
$f_{\rm NL}$ which contributes to the total power spectrum as $P_\zeta = P_{\zeta_g} + P_{\zeta_g}^{\rm NG}$, where $P_{\zeta_g}^{\rm NG} \sim f_{\rm NL}^2 A_\textrm{PBH}^2$.
In the UV  ($k \gg k_0$) and IR ($k\ll k_0$) regimes, one can follow the similar arguments as for the induced GW spectrum in Section~\ref{sec:IGWproduction} to find that
\begin{align}
P_{\zeta_g}^{\rm NG}(k \gg k_0) \sim  \left(\frac{k}{k_0}\right)^{-n_{\textrm{UV}}}, \quad
P_{\zeta_g}^{\rm NG}(k \ll k_0) \sim  \left(\frac{k}{k_0}\right)^{3}. 
\end{align}
Note that there is no logarithmic running in the IR regime since the curvature perturbations are always computed on superhorizon scales. More detailed calculations for the non-Gaussian power spectrum of USR inflation can be found in Refs.~ \cite{Atal:2021jyo,Adshead:2021hnm,Ragavendra:2020sop,Ragavendra:2021qdu}.

%It was shown that peak and UV tails of the GW spectrum induced by the broken power-law curvature
%spectrum are well approximated by
%\begin{align}
 %   \mathcal{P}_{\mathcal{R}}^{\textrm{NG}}(k\gg k_0)\approx \frac{36}{25} f_{\textrm{NL}}^2 A_\textrm{PBH}^2\left(\frac{1}{n_{\textrm{UV}}}+\frac{1}{n_{\textrm{IR}}}\right)\left(\frac{k}{k_0}\right)^{-n_{\textrm{UV}}},
%\end{align}
%where in our case, $n_\textrm{IR}=4$ and $n_\textrm{UV}=-2\delta_2-6$ and the IR tail (where $n_\textrm{IR}=3/2$) is in general, approximated by  
%\begin{align}
 %   \mathcal{P}_{\mathcal{R}}^{\textrm{NG}}(k\ll k_0)\approx 2 f_{\textrm{NL}}^2 A_\textrm{PBH}^2\left(\frac{1}{2 n_{\textrm{UV}}+3}+\frac{1}{3n_{\textrm{IR}}+3}\right)\left(\frac{k}{k_0}\right)^3.
%\end{align}
For the single-field inflation scenarios considered in this work, the spectral tilt can be related to the non-Gaussianity parameter by $n_{\textrm{UV}}=\frac{5}{12} f_{\textrm{NL}}$ \cite{Atal:2018neu} which translates to the USR parameter via
\begin{align}
    f_{\textrm{NL}}=-\frac{5}{12}(6+2\delta_2).
\end{align}
Even when exaggerated scenarios are considered, such as where $f_{\textrm{NL}}\simeq 2$ in Ref.~\cite{Atal:2021jyo}, very little correction to the total curvature perturbations or to the induced GW background was realised. Our maximal value for non-Gaussianity corresponds to $f_{\textrm{NL}}=5/12$ from the steepest spectral tilt of $\delta_2=-3.5$ for the cosmic coincidence, and thus we can safely conclude that non-Gaussian corrections to $\Omega_{\textrm{IGW}}$ are subdominant in all cases of interest here as well.
%}

\subsubsection{Gravitational waves from binary mergers}\label{Sec_binary_merger}
In Section~\ref{sec:IGWproduction}, we focused on the GWs induced by large primordial density fluctuations during inflation. These large fluctuations
collapse to form PBHs if their RMS amplitude exceeds a threshold value. However, once PBHs have formed, they may also behave as source for other GWs independent from those produced during inflation. The other GW counterparts could be due to PBH mergers and graviton emission by Hawking evaporation \cite{Kohri:2018awv}. We note that if PBHs are lighter than $10^{15}$ g they have already evaporated by today and no mergers can be resolved. PBHs heavier than $10^{15}$ g have not yet evaporated and their binaries could be merging in the nearby universe. We avoid detailed discussion of signals from mergers in this work due to the large astrophysical uncertainties and strong model dependence associated with such predictions.  However, there do exist some basic approximations that provide very crude order of magnitude estimates of the frequency at which we expect the GWs from the mergers of PBH binaries to show up at the Innermost Stable Circular Orbit \cite{Maggiore:2007ulw}
\begin{align}
    f_{\textrm{GW,max}}\approx 2 f_\textrm{ISCO} \approx 4.4\,\textrm{kHz}\left(\frac{M_\odot}{M}\right), 
\end{align}
where M is the total mass of the binary merger. If we consider a monochromatic PBH mass function then it follows that $M=2M_{\textrm{PBH}}$. Then we may compute the approximate frequency in today's universe $f_{\textrm{GW,max,0}}=\frac{f_{\textrm{GW,max}}}{1+z}$, where $z$ denotes redshift. Hence for the PBH mergers considered in Section~\ref{Sec_PBH_DM} of mass $\simeq 10^{-12} M_{\odot}$, we get very high frequency GWs $\gtrsim 10^{12}$ Hz that are outside the scope of even the most futuristic experimental projections.

\section{Conclusion}\label{Sec_conclusion}

The sharp deceleration of the slow-roll dynamics on small scales into a transient ultra-slow-roll (USR) phase is a generic mechanism to enhance the primordial power spectrum for primordial black hole (PBH) formation in single-field inflation. If PBHs indeed play an important role as dark matter (DM), the cosmic coincidence problem along with the energy densities of the universe today might hint at a correlated origin for baryons and DM. In this work, we have explored and confirmed the viability of baryogenesis, based on the Affleck-Dine (AD) mechanism, with modified initial conditions driven by a generic USR transition for PBH formation. Our results include the constant-mass AD field as a special case considered in Refs.~\cite{Wu:2021gtd,Wu:2021mwy}.

%Baryogenesis driven by the USR transition during inflation significantly extends the viable parameter space for the initial conditions from preceding conclusions. Scalar fields with initial VEVs much smaller than the Hubble scale of inflation can create sufficient baryon asymmetry as long as they were dynamically excited during inflation. This is to say, the AD mechanism of baryogenesis can be realized with a minimal Kahlar potential or without the assumption of supersymmetry. Albeit the formalism presented in this work has been focused on the constant-mass assumption for the AD field in the general USR region ($-4<\delta_2 \simeq-3$) with a bound \textcolor{red}{$0 < m_\pm/H_\ast < 3/2$} on the effective masses, we found successful baryogenesis across this whole region with the requirement on the rate $\Delta N_c\simeq[0.5,3]$. \textcolor{red}{We have therefore considered coherent production of scalar motion away from the USR limit and the modified late-time behavior for $m_\pm/H_\ast > 3/2$}

In the generalised region $\delta_2 < -3$ away from the standard USR scenario, we find asymptotically constant behavior of the final baryon asymmetry $Y_B$ towards the large USR duration limit ($N_\ast \gg 1$) irrespective of the rate $\delta_2$. We find this to be generic feature of the model and this is directly connected to the cosmic coincidence:
 for PBHs to occupy a significant fraction or saturate the DM density today (or equivalently $f_{\rm PBH} = \Omega_{\textrm{PBHeq}}/\Omega_{\rm CDMeq} \lesssim 1$), the value of $N_\ast = \Delta N$ must be precisely fixed. The allowed parameter space of $N_\ast$ for PBH DM lies well within the constant plateau region for the correct baryon asymmetry, ensuring the specific ratio $\Omega_{\rm CDMeq}/\Omega_{\rm Beq} \approx 5$ within statistical uncertainties of the PBH abundance \cite{Wu:2021mwy}. 
 However, the constant plateau for the correct baryon asymmetry narrows with the decrease of the USR rate $\delta_2$ from $-3$, and thus PBH as all DM can be satisfied for $\Delta N\in[1.7,2.6]$ in the corresponding range $\delta_2\in[-3.5,-3.1]$. This is the condition for the presented scenario to provide a clear indication to the cosmic coincidence problem.
 
 The ultralight asteroid-mass window $ 10^{-16}\lesssim M_{\textrm{PBH}}/M_\odot \lesssim 10^{-11}$ for PBH DM corresponds to the choice of pivot scales $k_0\in [10^{12}, 3\times 10^{14}]\, \textrm{Mpc}^{-1}$ for the USR template. As the most promising observational consequences of the general USR inflation, we have computed the induced gravitational wave (GW) background for $f_{\rm PBH} = 1$ and found that it has 
 a peak frequency $10^{-3} < f_{\textrm{peak}} < 1 $ Hz with 
 a maximum spectral amplitude around $h^2\Omega_{\textrm{IGW0}}(f)\simeq 1\times10^{-8}$. The frequency scaling of the $\Omega_{\textrm{IGW}} \sim f^{-2n_{\rm UV}}$ spectrum in the limit of $k \gg k_0$ (namely the UV tail) is controlled by the choice of the USR rate $\delta_2$ (as $n_{\rm UV} = -2\delta_2 - 6$), where the current LIGO and Virgo constraints on the stochastic GW background \cite{KAGRA:2021kbb}
($\gtrsim 10^{-8}$ in the vicinity of 25 Hz) is about one order of magnitude higher than the highest UV tail for $\delta_2 = -3.1$. For PBHs comprising all DM from USR inflation in the mass window of interest, the IR tail of $\Omega_{\textrm{IGW0}}$ must be measured by LISA, TianQin or DECIGO. The UV tail of the induced GWs would be tested by future experiments such as LISA, Advanced LIGO and Virgo (beyond O3), the Einstein Telescope (ET) and DECIGO, regardless of whether it is compatible with the cosmic coincidence scenario considered or not.
 % could already exclude the ``cosmic coincidence'' range ($n_{\rm UV}\in[0.2, 1]$ with $\delta_2\in[-3.5,-3.1]$) for $k_0 \gtrsim 5\times 10^{13}$ Mpc$^{-1}$ (which translates to PBH masses $M_{\rm PBH} \lesssim 3.6\times 10^{-15} M_\odot$). UV tails compatible with the ``cosmic coincidence'' for pivot scales $k_0\in [10^{12}, 5\times 10^{13}]\, \textrm{Mpc}^{-1}$ (which are associated with $3.6\times 10^{-15} - 10^{-11} M_\odot$ PBH DM) could be tested by future experiments such as LISA, Advanced LIGO and Virgo (beyond O3), the Einstein Telescope and DECIGO.

%  The ultralight asteroid-mass window $M_{\textrm{PBH}}\simeq 10^{-15}\textrm{-}10^{-11} M_\odot$ for PBH dark matter, one of the fundamental assumptions in this work resulting from choice of pivot scale $k_0\in [10^{12} - 10^{14}]\, \textrm{Mpc}^{-1}$, could be tested by near future astrophysical experiments. 
 %The cosmic coincidence is conservatively found to occur in the region $-3.5\leq\delta_2\leq -3.1$.

%This is observationally interesting for experiments such as LISA, advanced LIGO, the Einstein Telescope and DECIGO. 

Further more, we find that non-Gaussianity would have subdominant effects on the total curvature power spectrum and GW background since $f_{\textrm{NL}}=-\frac{5}{12}(6+2\delta_2)$ is not large enough to source notieceable corrections \cite{Atal:2021jyo}.  We also conclude that gravitational waves resulting from binary mergers would be at frequencies $f_{\textrm{peak}}\geq 10^{12}$Hz which would be far too high for observation even in the most optimistic future experimental scenarios.
Finally, we remark that even though the USR transition provides a striking solution to the cosmic coincidence, the fine-tuning of inflationary parameters to properly realise PBH DM inevitably remains.

\section*{Acknowledgements} 
%We would like to thank Jacopo Fumagalli and Yoann Genolini for helpful discussions.
We would like to thank Guillem Domènech, Jacopo Fumagalli, Gabriele Franciolini and Yoann Genolini for helpful discussions and feedback.
The project has received funding from the European Union’s Horizon 2020 research and innovation programme under grant agreement No 101002846 (ERC CoG ``CosmoChart'') as well as support from the Initiative Physique des Infinis (IPI), a research training program of the Idex SUPER at Sorbonne Université.

\appendix

\section{The power spectrum of induced tensor perturbations}\label{Appd_tensor_spectrum}
We provide the complete equations for the computation of the tensor power spectrum $P_h$ used in Eq.~\eqref{eq:GWspectrum}. 
Let us begin with the definition of linear perturbations in the conformal Newtonian gauge for the metric of the form
\begin{align}
ds^2 = -a^2(\eta)(1+2\Psi) d\eta^2 +  a^2(\eta) \left[(1-2\Phi)\delta_{ij} + \frac{1}{2} h_{ij} \right] dx^i dx^j, 
\end{align}
where $\Psi$ is the Newtonian potential and $\Phi$ is the curvature potential. We define
\begin{align}
h_{ij}(\vec{x},\eta) = \int d^3k (2\pi)^{-3/2} [e_{ij}^+(\vec{k})h_+(\vec{k},\eta) + e_{ij}^\times (\vec{k})h_\times(\vec{k},\eta)] e^{i\vec{k}\cdot\vec{x}},
\end{align}
which is the linear tensor perturbation including the two polarisation modes. The transverse-traceless polarisation tensors are
\begin{align}
    e_{ij}^+(\vec{k}) &= \frac{[e_i^1(\vec{k})e_j^1(\vec{k}) - e_i^2(\vec{k})e_j^2(\vec{k})]}{\sqrt{2}},& 
e_{ij}^\times(\vec{k}) = \frac{[e^1_i(\vec{k}) e^2_j(\vec{k}) + e^2_i(\vec{k}) e^1_j(\vec{k})]}{\sqrt{2}},
\end{align}
which are expressed in terms of orthonormal basis vectors $\textbf{e}^1$ and $\textbf{e}^2$ orthogonal to $\vec{k}$.

Keeping the tensor perturbation at linear order and the linear scalar perturbations up to second order, one can obtain the equation of motion for each polarisation $h_\lambda$ from the Einstein equation as
\begin{align}\label{eom_h}
h_\lambda^{\prime\prime}(\vec{k},\eta) + 2\mathcal{H} h_\lambda^\prime (\vec{k},\eta)+ k^2 h_\lambda(\vec{k},\eta) = 4 S_\lambda (\vec{k},\eta),
\end{align} 
where second-order perturbations are projected away in the transverse-traceless decomposition \cite{Baumann:2007zm} and we have neglected the anisotropic stress in the energy momentum tensor so $\Psi = \Phi$, and thus
\begin{align}
S_\lambda(\vec{k},\eta) =& \int \frac{d^3q}{(2\pi)^{3/2}} e_{ij}^\lambda(\vec{k}) q^iq^j \psi_{\vec{p}}\psi_{\vec{p}} f(p,q,\eta), \\
f(p,q,\eta) =& 2T(p\eta)T(q\eta) + \frac{4}{3(1+w)} 
\left[\frac{T^\prime(p\eta)}{\mathcal{H}} + T(p\eta)\right] \left[\frac{T^\prime(q\eta)}{\mathcal{H}} + T(q\eta)\right],
\end{align}
where $\mathcal{H} = aH = \frac{2}{(1+3w)\eta}$ and $w$ is the equation of state of the universe. $\vec{k} = \vec{p} +\vec{q}$ with $p\equiv \vert\vec{p}\vert$, $q\equiv \vert\vec{q}\vert$ are the two internal momenta. The time evolution of the scalar potential is described by $\Psi_{\vec{k}}(\eta) = T(k\eta)\psi_{\vec{k}}$ with respect to the primordial value $\psi_{\vec{k}}$, where the transfer function in the radiation dominated universe is given by \eqref{def_window_transfer}. The primordial Newtonian potential $\psi_{\vec{k}}$ well outside the horizon is related to the (gauge-invariant) curvature perturbation $\zeta$ as $\psi_{\vec{k}} = \frac{3+3w}{5+3w}\zeta$, or namely
\begin{align}
\left\langle \psi_{\vec{k}}\psi_{\vec{K}}\right\rangle = \delta^{(3)}\left(\vec{k} +\vec{K}\right) \frac{2\pi^2}{k^3} \left(\frac{3+3w}{5+3w}\right)^2 P_\zeta(k).
\end{align}
This is where parameters of the inflationary scenario given by \eqref{def_USR_parameters} enters the power spectrum of the tensor perturbation. 

Solving the equation of motion \eqref{eom_h} by virtue of the Green's function method \eqref{sol_h_Green}, we can compute the total power spectrum of the tensor perturbation as
\begin{align}\label{def_P_h}
\delta^{(3)}(\vec{k}+\vec{K}) P_h(k, \eta) 
=& \frac{k^3}{2\pi^2} \sum_\lambda^{+,\times} \left\langle h_\lambda(\vec{k},\eta)h_\lambda(\vec{K},\eta)\right\rangle, \nonumber\\
=& \frac{k^3}{2\pi^2} \int^\eta d\eta_1 G_{\vec{k}}(\eta ;\eta_1) \frac{a(\eta_1)}{a(\eta)}
 \int^\eta d\eta_2 G_{\vec{K}}(\eta;\eta_2)   \frac{a(\eta_2)}{a(\eta)} \nonumber\\
 &\qquad\times \sum_\lambda^{+,\times} \left\langle S_\lambda(\vec{k},\eta_1)S_\lambda(\vec{K},\eta_2)\right\rangle.
\end{align}
Here, it is convenient to use the dimensionless variables $u \equiv p/k$, $v\equiv q/k$ and $z \equiv k\eta$ to rewrite the tensor spectrum as
\begin{align}\label{eq:P_h_uv_variable}
P_h(k, z) &= 4\int_{0}^{\infty} dv\int_{\vert 1-v \vert}^{1+v} du 
\left[\frac{v}{u} - \frac{(1-u^2+v^2)}{4uv}\right]^2 I^2(u,v,z) P_\zeta(ku) P_\zeta(kv), \\
I(u,v,z) &= \frac{9(1+w)^2}{(5+3w)^2}\int_{0}^z dz_1 \frac{a(z_1)}{a(z)} kG_{\vec{k}} (z,z_1) f(u,v,z), 
\end{align}
where our definition of $I(u,v,z)$ coincide with that defined in Ref.~\cite{Kohri:2018awv}. Note that the projection of momentum under polarisation tensors can be found in the Appendix B of Ref.~\cite{Atal:2021jyo}, where 
\begin{align}
 (e_{ij}^+ q_iq_j)^2 + (e_{ij}^\times q_iq_j)^2 = k^4v^4\left[1-\frac{(1-u^2+v^2)^2}{(2v)^2}\right]^2.   
\end{align}
For numerical evaluation, we adopt new variables $t = u+v-1$, $s =u-v$ introduced in Ref.~\cite{Kohri:2018awv}, where $u=\frac{t+s+1}{2}$, $v=\frac{t-s+1}{2}$ and the tensor spectrum now reads
\begin{align}
\label{eq:powerspectrum}
P_h(k, z) =& 2\int_{0}^{\infty} dt \int_{-1}^{1} ds \left[ \frac{t (2 + t) (s^2 - 1)}{(1 - s + t) (1 + s + t)}\right]^2 \\\nonumber
%P_\zeta\left(k u\right) P_\zeta\left(k v\right) I^2(s,t,x),
&\quad \times P_\zeta\left(\frac{k (t+s+1)}{2}\right) P_\zeta\left(\frac{k (t-s+1)}{2}\right) I_{\rm RD}^2(s,t, z).
\end{align}
% and \textcolor{red}{dimensionless} parameter $x=k\eta$. 
In the late-time limit of the radiation dominated universe i.e. where $\eta\rightarrow\infty$ and $z \gg 1$, we have the oscillation averaged result from Ref.~\cite{Kohri:2018awv} as
\begin{align}
\label{eq:GWtf}
&\overline{I^2_{\textrm{RD}}(s,t, k\eta \to \infty)} =  \frac{288 (-5 + s^2 + t (2 + t))^2}{z^2(1 - s + t)^6 (1 + s + t)^6} \times
\\
&\qquad\Bigg\{ \frac{\pi^2}{4} \left(-5 + s^2 + t (2 + t)\right)^2 \Theta\left(t-(\sqrt{3}-1)\right) +
\nonumber \\\nonumber
&\qquad \left[-(t - s + 1) (t + s + 1) + 
\frac{1}{2} (-5 + s^2 + t (2 + t)) \ln\left|\frac{(-2 + t (2 + t))}{3 - s^2}\right|\right]^2\Bigg\},
\end{align}
where $\Theta$ is the usual Heaviside theta function. Hence the averaged analytical transfer function during radiation domination \eqref{eq:GWtf} above can be substituted into Eq.~\eqref{eq:powerspectrum}. The resulting integral will yield the oscillation averaged power spectrum $\overline{P_h(k,\eta)}$ which can be substituted into Eq.~\eqref{eq:GWspectrum} and the dimensionless gravitational wave background to be compared with experimental limits or signals can be determined. We have verified the calculation by direct comparison with the scale invariant power spectrum normalised to unity ($A_\zeta = 1$), which yields a dimensionless gravitational wave spectrum of $\Omega_{\textrm{IGW}}/A_\zeta^2 = 0.822$ as expected in Ref.~\cite{Kohri:2018awv}.

\bibliographystyle{JHEP} 
\bibliography{refs_inf}
\end{document}